\documentclass[preprint]{aastex62}

\newcommand{\bnabla}{\mbox{\boldmath$\nabla$}}
\newcommand{\kB}{k_{\rm B}}
\usepackage{amsmath}
\usepackage{epstopdf}
\usepackage{graphics}
\newcommand{\nn}{\mbox{} \nonumber \\ \mbox{}} 
\usepackage{comment} 

\graphicspath{{./}{figures/}}

\accepted{for Publication in the Astrophysical Journal}

\begin{document}

\title[Limits on Protoplanet Growth]{Limits on Protoplanet Growth by Accretion of Small Solids}

\correspondingauthor{M. Ali-Dib, C. Thompson}
\email{malidib@astro.umontreal.ca, thompson@cita.utoronto.ca}

\author{Mohamad Ali-Dib}
\affiliation{Institut de recherche sur les exoplan\`etes, D\'epartement de physique, Universit\'e de Montr\'eal. 2900 boul. \'Edouard-Montpetit, Montr\'eal, Quebec, H3T 1J4, Canada}
\affil{Centre for Planetary Sciences, Department of Physical \& Environmental Sciences, University of Toronto Scarborough,
Toronto, ON M1C 1A4, Canada}
\affiliation{Canadian Institute for Theoretical Astrophysics, 60 St. George St, University of Toronto, Toronto, ON M5S 3H8, Canada}

\author{Christopher Thompson}
\affiliation{Canadian Institute for Theoretical Astrophysics, 60 St. George St, University of Toronto, Toronto, ON M5S 3H8, Canada}


\begin{abstract}
This paper identifies constraints on the growth of a small planetary core ($0.3\,M_\oplus$)
that accretes millimeter-sized pebbles from a gaseous disk. We construct time-dependent spherical envelope models that
capture physical processes  
not included in existing global hydrodynamic simulations, including particle size evolution,
dust transport, and realistic gas equations of state. We assume a low enough disk density that
pebbles are marginally coupled to the gas and are trapped efficiently near the core Bondi radius.
Pebbles then drift rapidly enough to experience erosion by sandblasting,
mutual collisions, and sublimation of water ice.  We find that pebble fragmentation is more efficient than dust re-sticking.
Therefore the large pebble accretion rate $\dot M_p$ needed
to build a core of mass $> M_\oplus$ leads to a high envelope metallicity and grain opacity.  
Above $\dot M_p\sim 1\times 10^{-7}\,M_\oplus$ yr$^{-1}$, and without other luminosity sources,
convective motions expand near the Bondi radius.  
The warm, dusty, and turbulent envelope buffers the inward drift of pebble debris:
given a turbulent concentration factor $f_{\rm turb} \gtrsim 1$ near the 
lower convective boundary, the core growth rate is limited to
$1\times 10^{-7}\,f_{\rm turb}\,M_\oplus$~yr$^{-1}$ and the e-folding time $3/f_{\rm turb}$ Myr.
The remainder of the solid debris is expelled as highly processed  silicates.
Pebble ice never reaches the core, and the envelope contains comparable amounts of H$_2$/He and metals. 
We interpret our results using simpler steady models and semi-analytical estimates.  Future 
global simulations incorporating the processes modelled here are needed to understand the influence
of rotation and vertical disk structure.
\end{abstract}


\vfil

\section{Introduction}

The accretion of small, collisional ``pebbles'' has been suggested as an efficient channel
for planet growth, in the first instance in order to explain the assembly
of Uranus and Neptune in our solar system.  Pebble accretion may
be enhanced either by collisional cooling in a thin, pebble-dominated disk \citep{goldreich2004} 
or by gas drag \citep{ormelklahr,lamborg}.  Here we focus on pebble
accretion from a gaseous protoplanetary disk (PPD), examining how collisional effects
may limit core growth from pebble debris.   

A key constraint on this accretion channel is that the mm-cm sized pebbles must only be mildly coupled
to the gas near the planetary Bondi radius, as defined by the temperature of the ambient gas disk: 
if the coupling is tight then pebble inflow is limited by the residency of
gas outside the Bondi radius.  Since the flow speed near the Bondi radius is typically a modest fraction of the
sound speed (e.g. \citealt{popovas}),
accreting pebbles must drift at relatively high speeds ($\sim 100$ m s$^{-1}$) with respect to the gas.
Pebbles formed by the adhesion of micron-sized grains through van der Waals forces are fragile,
and binary collisions at such speeds are destructive \citep{dustnew1,dustnew2,dustnew3}.  These collisions 
occur with significant optical depth if the pebble accretion rate is high enough to grow a core 
above an Earth mass over an interval of 1 Myr.

Even faster abrasion of pebbles occurs, as we show, through collisions with dust grains embedded
in the gas (see \citealt{jacquet} in the context of solid-gas shocks).  Small target grains
have a much longer residency time in the envelope than do the pebbles, meaning that their abundance
is not fixed and can substantially exceed the abundance in the ambient disk.  Pebble abrasion
combined with mutual fragmenting collisions between the small particles causes an exponential rise
in the envelope dust abundance.

Icy pebbles are also destroyed by rapid sublimation as the temperature
rises above 100~K.  Pebble destruction
inhibits the direct incorporation of water or silicates into the 
core, but does allow enrichment in
dust and the extraction of heat from the gaseous component through ice sublimation
\citep{stevenson84, hori2011, chambers2017}.  The net result is that gas around
the Bondi radius can become sufficiently dusty to suppress
radiative transport, well outside the inner hydrostatic gaseous envelope
where pebble destruction has previously been considered
 \citep{pollack1986, podolak, vent, Brouwers}.  

The retention of gas and solids depends on efficient radiative cooling (e.g. \citealt{lee}).  When
cooling is weak, \cite{ormelcy} and \cite{alibert} showed using three-dimensional
isothermal simulations that the accreted envelope surrounding a core behaves like an open system,
with the envelope material continually being recycled back into the gas disk.  As a result, the 
accretion of pebbles from the neighboring disk into a bound, quasi-spherical envelope 
(concentrated inside the Bondi radius of the planetary core) may be intrinsically inefficient 
\citep{kuwahara}.  \cite{lega} included heating by planetesimal
accretion onto the core surface and energy exchange between the gas and diffusing
radiation.  \cite{kurokawa} also considered the case where gas accreted close
to the core is able to cool radiatively, but ignored the potentially strong heating effects
associated with the accretion of a modest number of large planetesimals.
The refluxing of envelope material away from higher-mass cores, driven by 
planetesimal accretion, has been considered by \cite{rafikov2006}  
in an analytic spherical envelope model with parameterized opacity, but the effects of dust production and
ice sublimation were not included.

Here we demonstrate (i) an even stronger suppression of direct pebble accretion onto the core than has
previously been obtained; and 
(ii) a limitation on envelope growth due to the reflux of gas and dust by convective motions back into the disk.
A key result is that fragmentation can become a runaway process, so that the density of small particles
  settling toward the core is strongly enhanced as their size and inward drift speed is reduced.  Previous models of dust
  deposition in the envelope typically include growth by adhesion but not collisional fragmentation, and imply
  inward drift speeds exceeding $\sim 10^2$ m s$^{-1}$ \citep{mordaop, ormel}.
  
  We find that the limitation on core accretion is strongest for the water ice carried by pebbles.
  In addition, the H$_2$/He
  mass accumulated in the atmosphere is strongly tied to the onset of convection near the Bondi radius.
We develop a time-dependent model of the structure of the gas
envelope accreted from a PPD that includes the feedback of pebble destruction on radiative opacity and the possibility of convection
near and beyond the Bondi radius.  The fiducial core mass is $0.3\,M_\oplus$,
which is significantly heavier than the bound planetesimals
which have been found to collapse directly via the streaming instability in pebble-rich gas \citep{jonlam}.
Cores of around this fiducial mass were found in hydrodynamic simulations to accrete pebbles vigorously \citep{lamborg, lamb1}.

The capture rate of pebbles by a planetary core could cover a range of values, and
a goal of this study is to determine the critical rate above which the envelope is
heated rapidly enough, and its opacity is high enough, that much of the pebble debris
is returned to the disk.  The limiting accretion rate $\dot M_p$ so obtained can
  be compared with the rate needed to grow the core above a mass $M_c \sim (1-10)M_\oplus$
  in the lifetime of a PPD.
In a variant on our default luminosity model, we 
also consider an additional source of heating by the accretion of large 
planetesimals that easily penetrate the convective envelope;  this
can compete with (or dominate) the gravitational heating by infalling pebble debris.

 An important consideration is the rate at which pebble debris circulating in
the lower envelope will be incorporated into the core.  \cite{popovas, popovas19} have calculated
pebble trajectories in a three-dimensional simulation of a core embedded in a shearing
gas disk, finding rapid downward advection of pebbles by convective flows.  Pebbles
reaching the computational core boundary are allowed to be accreted, releasing gravitational
energy and driving convection around the core.  Although we also find that turbulent
diffusion of grains through the convective layer is faster than secular drift in the central
gravitational field, particles drifting at the silicate fragmentation speed $\sim 1$ m s$^{-1}$ 
are very tightly coupled to the gas.  In fact, the simulations of \cite{popovas, popovas19} do
not resolve the dynamics of small particles with short stopping times near the core boundary.
We evaluate the core growth rate in terms of the drift rate of embedded grains 
across the lower convective-radiative boundary, which is proportional to the dust concentration.
The turbulent intensity decreases toward the convective boundary, meaning that this
concentration may be enhanced with respect to the interior of the convection zone, by a 
factor $f_{\rm turb}$.  The core accretion rate is evaluated for $f_{\rm turb} = 1$, but
the possibility of an enhancement due to turbulent pumping should be kept in mind.


The plan of this paper is as follows.  Our treatment of the
interactions of solid particles with gas, and with each other,
is described in Section \ref{s:estimates}.   Section \ref{s:expand} presents
a simple analytical steady-state model showing how convection expands within a hydrostatic
envelope that is progressively loaded with small particles.  Section \ref{s:model} describes our one-dimensional (spherical)
hydrostatic numerical model of the gas envelope, including heat exchange
between sublimating solids and gas and the convective transport of embedded particles calculated in the mixing-length approximation.
Results are presented in Section \ref{s:results} for
various pebble accretion rates in the $\dot M_p = 10^{-7}-10^{-6}\,M_\oplus$ yr$^{-1}$ range (corresponding 
to a growth times $\sim 10^6-10^7$ yr to reach $10\,M_\oplus$ starting from $0.3\,M_\oplus$),
and accretion rates $\dot M_{\rm plan} = 0$, $10^{-7}\,M_\oplus$ yr$^{-1}$ of larger planetesimals.
The effect of pebble composition (icy vs. dry)
is also considered.  The implications of our work and some
outstanding issues are summarized in Section \ref{conc}.  The
Appendix details a simple analytic model which is consistent
with the detailed numerical results.

\section{Interactions of Solid Particles}\label{s:estimates}

Destructive processes acting on pebbles are the main subject of this section.  We begin
by summarizing why fast pebble drift is characteristic of efficient pebble accretion.
A high pebble accretion rate (needed to build a substantial core over a Myr interval)
is connected to a significant optical depth to destructive binary collisions between pebbles. 
We explain the relative importance of sandblasting by small grains embedded in the gas,
and show that dust production is not limited by resticking of grains before the dust/gas
ratio reaches of the order unity.  

Next we quantify our treatment of pebble destruction in the numerical model, 
defining a self-consistent one-size model for pebbles embedded in a turbulent, convective
atmosphere.   Although our default envelope model does not include the effect of silicate
dust sublimation on core accretion, we present additional arguments supporting the
suppression of silicate `rainout' in the inner envelope, 
as a consequence of the large atmospheric scale height and silicate mass fraction.

In what follows, we use `pebble' to denote 0.1-1 cm sized objects, and `planetesimal'
to denote 1-10 km sized objects.  We focus on the regime $\dot M_p \gg \dot M_{\rm plan}$:
pebbles are the main {\it potential} source of planetary material, but because the pebbles
disintegrate in the outer envelope, planetesimals can supplement the heat flux
through the envelope.

\subsection{Some Introductory Estimates}

The fiducial core mass is\footnote{This is also comparable to the planetesimal 
isolation mass near 5 AU in the minimum-mass solar nebula of \cite{hayashi} \citep{goldreich2004b}.}
$0.3\,M_\oplus$, which is light enough that gas pressure gradient forces play a significant role in determining
the gas profile within the core's tidal radius: in other words, the Bondi radius is smaller than
the Hill radius\footnote{Throughout this paper, we use the shorthand $X = X_n\times 10^n$, where quantity $X$ is 
presented in c.g.s. units.  For example $\rho_{s,0}$ is the pebble density in units of $1 = 10^0$ 
g cm$^{-3}$.}
($R_{\rm H} = a_{\rm orb}(M_c/3M_\odot)^{1/3}$),
\begin{eqnarray}
    R_{\rm B} &=& {GM_c\over c_{g,\rm disk}^2} = 3.3\times 10^{10}\,T_{\rm disk,2}^{-1}\left({M_c\over 0.3~M_\oplus}\right)\quad {\rm cm};\nn
    {R_{\rm B}\over R_{\rm H}} &=& 0.33\,T_{\rm disk,2}^{-1}\left({M_c\over 0.3~M_\oplus}\right)^{2/3}\,
    \left({a_{\rm orb}\over {\rm AU}}\right)^{-1},
\end{eqnarray}
where $a_{\rm orb}$ is the orbital semi-major axis around a Solar-mass star
and $c_{g,\rm disk} = (k_{\rm B}T_{\rm disk}/\mu_g)^{1/2}$ is the isothermal gas sound
speed in the ambient disk of temperature $T_{\rm disk}$ and mean molecular weight $\mu_g$.
(In what follows, ``disk'' refers to the medium surrounding the core.) 
Such a core is just massive enough to sublimate silicate
grains embedded in a high-entropy H$_2$/He atmosphere, but its gravity is too weak
for the impacts of marginally-bound large planetesimals to convert significant silicate
mass to vapor.  

Efficient accretion of pebbles inside the Bondi radius implies an upper bound on the PPD
gas density $\rho_{g,\rm disk}$.  Writing the gas flow speed around the core outside the Bondi radius as 
$V_g = \varepsilon_V c_{g,\rm disk}$,
we require that the pebble drift speed $\Delta V$ with respect to the gas is 
$\Delta V \gtrsim V_g$ for the pebble to be trapped.  At some radius around $R_{\rm B}$, the gas
flow bifurcates from inflow-outflow to a hydrostatic, rotational motions (e.g. \citealt{kurokawa}). 
Near this bifurcation point, $\Delta V \sim \tau_{\rm stop}
GM_c/R_{\rm B}^2 \sim \tau_{\rm stop} c_g^4/GM_c$.  The radius $a_p$ of a trapped pebble is typically
smaller than the gas mean-free path, and so the stopping time
$\tau_{\rm stop} \sim \rho_s a_p/\rho_{g,\rm disk} c_{g,\rm disk}$,
where $\rho_s$ is the density of the pebble material.  Then the condition $\Delta V \gtrsim V_g$ 
implies
\begin{equation}\label{eq:rhodisk}
\rho_{g,\rm disk} \lesssim 1\times 10^{-11}\, {T_{\rm disk,2}a_{p,-1}\rho_{s,0}\over \varepsilon_{V,-1}}
\left({M_c\over 0.3~M_\oplus}\right)^{-1}\quad {\rm g~cm^{-3}}.
\end{equation}
This density is typical of the zone outside the water ice line (semi-major axis $\gtrsim 3$ AU) when about one Jupiter mass
of gas remains in the PPD.  For an orbit typical of an extra-Solar super-Earth (0.1~AU), 
it implies a low gas surface density $\sim 1$ g cm$^{-2}$ (corresponding to inefficient planetary migration
by planet-disk torques).

\begin{figure}
\begin{centering}
\epsscale{1.15}
\plotone{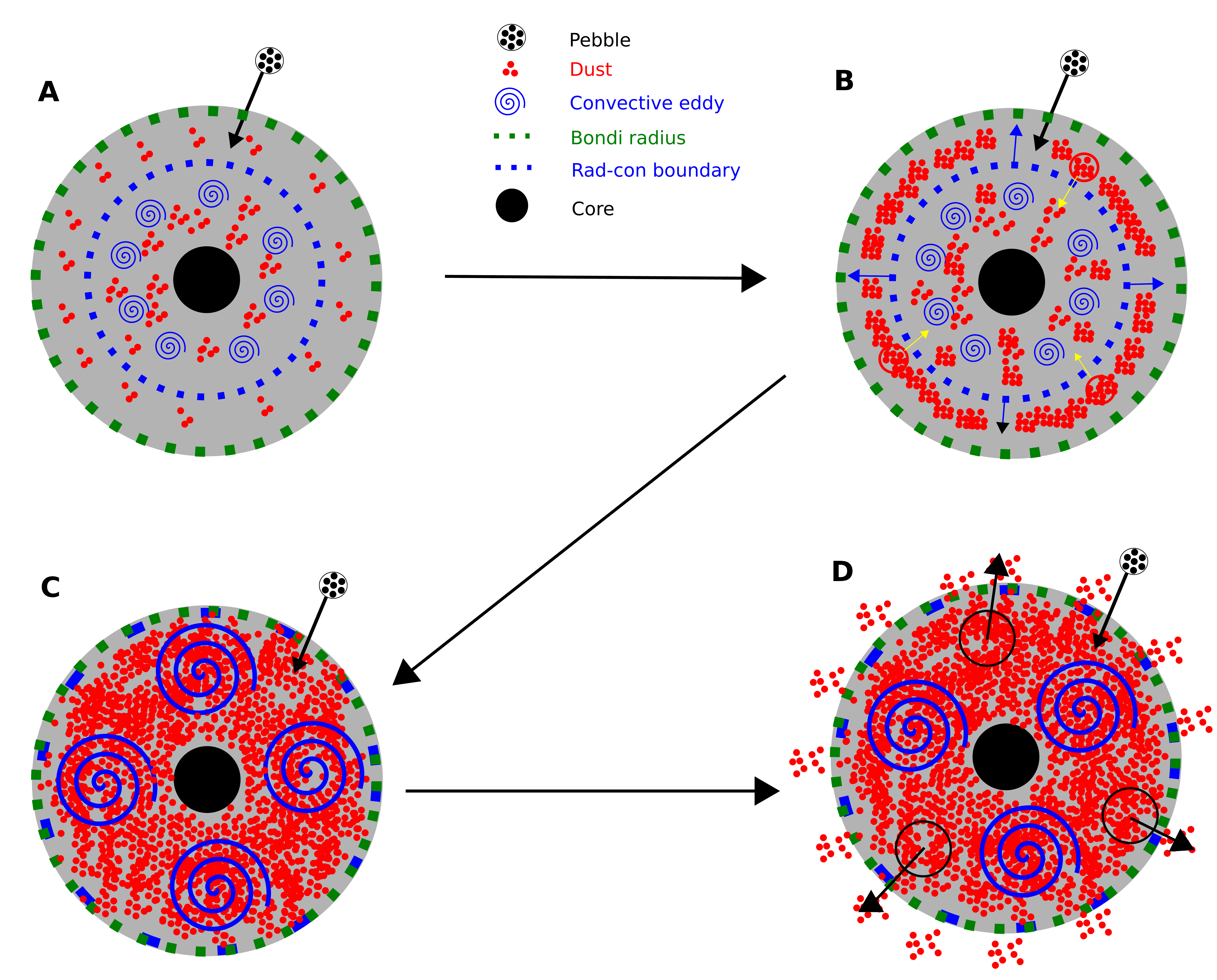}
   \caption{A schematic illustration of our numerical model, in the high-luminosity case where envelope growth
is stalled and a steady state is reached. In panel A, accreted pebbles are sandblasted into small dust in the outer
radiative layer of the hydrostatic gas envelope surrounding a condensed core that is embedded in a PPD.
Our radiative envelope is initiated with  ISM metallicity, hence the presence of ``seed'' grains.
In panel B, the rising dust abundance, coupled with the inward secular drift of the grains, increases the opacity and
pushes the radiative-convective boundary outward. While inward drift tends to remove dust from the radiative
zone, pebbles are continuously destroyed, replenishing the grains.  The grid points outside of the RCB will therefore
always contain dust.  In panel C, the runaway increase in dust abundance has pushed the radiative-convective boundary
out to the Bondi radius. In panel D, convection extends beyond the Bondi radius and effectively ejects 
dust-enriched gas back into the disk, stalling the growth of the envelope. This is the quasi-steady state in which a 
high-luminosity model remains stuck, where the continuous pebble accretion is counterbalanced by dust ejection.
Alternatively, a low-$\dot M_p$ envelope that sustains continued pebble and gas accretion does not progress beyond panel B.}
    \label{fig:drawing}
    \end{centering}
\end{figure}

Our model is schematically illustrated in Figure \ref{fig:drawing}.
Pebbles experience destructive collisions passing through dust-rich gas near the outer boundary of the bound envelope.
When solid particles are subject to a strong, transient acceleration, they drift differentially through the
gas with respect to each other, and especially with respect to small grains stuck in the gas.
High-speed collisions with dust grains embedded in the gas effectively sandblast away the pebble 
material (see \citealt{schrap1,schrap2} for experimental constraints and the application to PPDs
and \citealt{jacquet} for solid-gas shocks).  This destruction channel, which builds a strongly bimodal size 
distribution, is more effective than collisions between similarly-sized particles, which are the dominant 
process in a mildly turbulent PPD (e.g. \citealt{krijt}).  For a typical drift speed
$10^2$ m s$^{-1}$, a grain is expected to liberate $Y = 10-100$ times the grain mass from the pebble 
surface \citep{jacquet, schrap2}. 

When the seed dust density is small, binary collisions between pebbles provide an additional population
of small targets.  Around the Bondi radius, the pebble accretion rate
\begin{equation}\label{eq:dotMp}
\dot M_p \sim 4\pi r^2 \tau_{\rm stop} {GM_c\over r^2} \cdot n_p {4\pi\over 3}\rho_s a_p^3,
\end{equation}
where $n_p$ is the space density of pebbles.  We can write the accretion rate in terms of an accretion
time, $\dot M_p =  M_c/t_{{\rm acc},p}$.  Then the optical depth for binary collisions within
an outer radiative layer of the envelope ($\rho_g > \rho_{g,\rm disk}$, $T \simeq T_{\rm disk}$) is
\begin{equation}\label{eq:taupp}
\tau_{pp}(r \lesssim R_{\rm B}) \sim n_p 4\pi a_p^2 r \sim  2 \left({\rho_g\over \rho_{g,\rm disk}}\right) 
{\rho_{g,-11}\over T_2^{1/2} (a_{p,-1}\rho_{s,0})^2} \left({\dot M_p\over 10^{-6}\,M_\oplus~{\rm yr}^{-1}}\right)
\left({r\over R_{\rm B}}\right).
\end{equation}
Here we have restored a numerical factor in the gas drag law (Equation (\ref{eq:ts})). This expression
can also be expressed in terms of the gas flow speed $\varepsilon_V$ through Equation (\ref{eq:rhodisk}).
It applies in a situation preceding the build-up of small grains by sandblasting and
ice sublimation, in which case the outer envelope is nearly isothermal but the gas density
increases exponentially inward (Section \ref{s:struc}).
Binary pebble collisions are not included in our time-dependent envelope model due to the relative
effectiveness of ice sublimation and sandblasting.

Turning now to pebble-dust interactions, each infalling pebble intersects a column 
$X_d\rho_g GM_c/c_g^2 \sim X_d \rho_{g,-11}T_2^{-1}(M_c/M_\oplus)$ g cm$^{-2}$ of  
grains of mass fraction $X_d$.  Small grains have a long residency time in the gas,
and their density exponentiates within the bound envelope in response to the pebble ablation. 
We find that the mass fraction of particles smaller than the wavelength peak of the Planck function
($\lambda_{\rm max}(T_{\rm disk}) \sim  30\,\mu$m; Equation (\ref{eq:Q})) must exceed $X_d \sim 0.3$ before the outer envelope switches to a
convective state (Section \ref{s:2layer}; Figure \ref{fig:massratio}). As a result,
the column of intercepted grains is high enough that, 
combined with a large mass-loss multiplicity, each mm-cm sized pebble is destroyed already near the
outer boundary of the envelope.  Our numerical model, described in Section \ref{s:model}, self-consistently
calculates the grain density in response to the sourcing from grain-pebble collisions, and the loss by advection
out beyond the Bondi radius.  

Given the relatively high abundance of the target particles, one must also consider interactions between
them.  The particle size as limited by
binary fragmenting collisions is quite small.  Setting the radial drift speed through the static
gas equal to the fragmentation speed, $\Delta V(a_d) = \tau_{\rm stop} GM_c/r^2 = V_f$, one finds near the Bondi radius,
\begin{equation}\label{eq:admax}
a_{d,\rm frag}
\sim {9V_f GM_c \rho_{g,\rm B} \over 4\rho_s c_{g,\rm B}} = 4\;{V_{f,2}\rho_{g,\rm B,-11}\over T_{\rm B,2}^{3/2}}
 \left({M_c\over 0.3~M_\oplus}\right)\quad \mu{\rm m}.
\end{equation}
Secondary collisions between similarly-sized collision fragments larger than $a_{d,\rm frag}$
will also be destructive (e.g. \citealt{guttler10}), and have a significant optical depth when the inward
mass flow carried by the fragments rises to a fraction $\sim (a_d/a_p)^2$ of $\dot M_p$ 
(consider Equation (\ref{eq:taupp}) with $a_p$ replaced by $a_d \ll a_p$).
This means that the particle fragments with long residency times that are liberated by pebble-grain collisions
will, as their abundance rises, break down to a near-monomer size.  The pebble debris
is therefore an effective source of radiative opacity.
The interactions of even smaller (e.g. sub-micron sized) grains are considered in Section \ref{s:xdmax}.

Eventually, the evolution of grains in the inner bound gas envelope is determined by convectively forced collisions.
At high pebble accretion rates,
the convective energy flux is dominated by a small fraction of the pebble material that
settles onto the core,
with a corresponding accretion time $t_{{\rm acc},c} = M_c/\dot M_{\rm sett}(R_c)$ and
luminosity $GM_c^2/R_c t_{{\rm acc}, c}$. 
Then the convective Mach number reaches reaches 
\begin{equation}\label{eq:mach}
{V_{\rm con}\over c_g} \sim \left({GM_c^2\over R_c t_{{\rm acc},c}}\right)^{1/3} 
\left(4\pi R_{\rm B}^2 \rho_g c_g^3\right)^{-1/3} \sim 
0.3\,{T_2^{1/6}\over\rho_{g,-11}^{1/3} R_{c,9}^{1/3}} \left({t_{{\rm acc},c}\over {\rm 10~Myr}}\right)^{-1/3}
\end{equation}
at the Bondi radius.  
The turbulent acceleration of particles within these dynamic gas flows is high enough to prevent fragmented particles
from re-sticking, thereby maintaining a dense population of micron-sized grains.

\subsection{Pebble Dynamics and Destruction}

Solid particles in the envelope are divided into two components:  (i) large pebbles (of radius $a_p \sim$ mm), 
which are injected by accretion inside the Bondi radius;  and (ii) dust grains, 
which are produced mainly by `sandblasting' of pebbles during collisions with 
other grains embedded in the gas, and whose size deeper in the envelope is regulated by
binary collisions.

Pebbles are less tightly coupled to the gas than the smaller grains, and so drift more rapidly
with respect to the local gas flow.  Nonetheless, mm-sized pebbles are at least marginally coupled 
to the convective motions in the outer envelope, once these motions develop in response to the growing
radiative opacity.  The mean free path
between gas molecules is $\ell_g \equiv \mu_{\rm H}/\sigma_{\rm H} \rho_g \sim 10^2\rho_{g,-11}^{-1}$ cm
in the outer envelope.  Here $\sigma_{\rm H} \sim 3\times 10^{-15}$ cm$^{-2}$ is the mutual collision
cross section of H$_2$ molecules.  As a result, $\ell_g > a$ for both grains and pebbles, 
and the drag force is
\begin{equation}
 F_{\rm D}  = - {4\pi\over 3} \rho_g a^2 v_{\rm th}\, \Delta V\quad\quad (\rm Epstein),
\end{equation}
where $v_{\rm th} \sim 1.6 c_g = 1.6(kT_g/\mu_g)^{1/2}$ is the mean thermal speed of a gas molecule.  
In the inner envelope, fragmenting grains can have radii larger than $\ell_g$, whereas $\Delta V$ 
is much smaller than the gas sound speed.  Then the drag is in the Stokes regime,
\begin{equation}
    F_{\rm D} = - 6\pi \eta a  \Delta V\quad\quad ({\rm Stokes}),
\end{equation}
where $\eta \simeq (1/3)\rho_g v_{\rm th} \ell_g$ is the viscosity.  For the stopping time
$\tau_{\rm stop} = (4\pi/3)\rho_s a^3\Delta V/F_{\rm D}$ we use the interpolation formula
\begin{equation}\label{eq:ts}
\tau_{\rm stop} = {\rho_s a\over \rho_g c_g}\times {\rm max}\left({4\over 9}, {a\over\ell_g}\right).
\end{equation}

We take into account the erosion of pebbles by two effects:  sublimation and sandblasting.
Pebbles will sublimate as they drift inward into the hotter parts of the envelope.  Where the
water vapor pressure is below the saturation value, icy pebbles shrink at the rate
\citep{podolak}
\begin{equation}
\label{eq:pebdes1}
\frac{d a_p}{d t} = -0.63\left({\mu_{\rm H_2O} \over 2 \pi k T} \right)^{1/2} \frac{P_{\rm  H_2O}^{\rm sat}}{\rho_s}.
\end{equation}
Here, the numerical factor matches the measured sublimation rate at low pressure \citep{haynes}. 
In addition, as pebbles drift into the envelope, they will be sandblasted by small dust particles coupled to the gas.
This will result in the mass loss rate:
\begin{equation}
\label{eq:pebdes2}
{d\over dt}\left({4\pi\over 3}\rho_s a_p^3\right) = -Y(\Delta V) \times \pi a_p^2 \rho_d \Delta V.
\end{equation}
In the above $Y(\Delta V)$ is a yield factor controlling the pebble mass ejected per collision 
\begin{equation}
    Y ( \Delta V ) = 0.7\,\left( \frac {\Delta V} {V_f}  \right)^\alpha,
\end{equation}
where $V_f$ is the fragmentation velocity, and the coefficient and the index $\alpha \sim 1$ are
determined by laboratory experiments \citep{schrap2}.
In our atmosphere models $Y = O(10-30)$ in the outer pebble destruction zone.  
In practice, we find that sandblasting is the dominant destruction channel.

\subsection{Limitation on Very Small Grains in Outer Radiative Layer}\label{s:xdmax}

The accumulation of very small dust grains (of a radius smaller than Equation (\ref{eq:admax}))
may eventually be limited by mutual sticking.  Here we focus on an outer
radiative layer of the envelope.  
Sticking becomes important when the dust mass fraction $X_d$ reaches a level which we now estimate.
This is proportional to the pebble accretion rate $\dot M_p$ and also to the multiplication factor
$Y$ appearing in the mass loss formula (\ref{eq:pebdes2}).  Given that the grains are effectively stuck
in the gas, the net rate of change of the grain volume density is
\begin{equation}
\dot n_d = n_dn_p \cdot Y \pi a_p^2 \Delta V(a_p) - n_d^2 \cdot 4\pi a_d^2 \Delta V(a_d).
\end{equation}
Here both grains and pebbles are in the Epstein drag regime.
Relating the pebble volume density to the accretion rate via Equation (\ref{eq:dotMp}), we obtain
\begin{equation}
\dot n_d = {3Y n_d\dot M_p\over 16\pi r^2 \rho_s a_p}\left(1 - {X_d\over X_{d,\rm col}}\right),
\end{equation}
where for pebbles of icy composition,
\begin{equation}\label{eq:xdmax}
X_{d,\rm col} = {9 Y \dot M_p c_g\over 64\pi GM_c \rho_s a_p} = 0.47 {T_2^{1/2}\over a_{p,-1}}
\left({Y\over 10}\right)\left({t_{{\rm acc},p}\over 0.3~{\rm Myr}}\right)^{-1}
\end{equation}
and the pebble accretion time $t_{{\rm acc},p} \equiv M_c/\dot M_p$.
This dust mass fraction is comparable to the equilibrium metallicities observed
in our time-evolved envelope models, for a pebble accretion rate $\dot M_p = 10^{-6}\,M_\oplus$ yr$^{-1}$.

\subsection{Equilibrium Dust Size in the Convective Atmosphere}\label{s:dustsize}

Interior to the pebble erosion layer, we represent the size of the embedded dust particles
as being locally peaked at a value $a_d$ that is regulated by fragmentation and the
rate of mutual collisions.  At high accretion luminosities, dust grains in the upper envelope are stirred
rapidly by turbulence, and broken down to sizes smaller than (\ref{eq:admax}).
In the inner envelope, the collision speed between dust particles is
determined by the direct gravitational acceleration.  

First consider collisions between grains that are exposed to convectively driven turbulence.
Neglecting the effects of rotation, the eddy speed at a scale $\ell \ll r$ is $V_t \sim V_{\rm con} (\ell/r)^{1/3}$.
Small particles with stopping time $\tau_{\rm stop}$ decouple
from the convective eddies at a scale where $\ell/V_t \sim \tau_{\rm stop}$, and collide with each other at
a speed $\Delta V \sim V_t \sim V_{\rm con}(V_{\rm con}\tau_{\rm stop}/r)^{1/2}$.  This collision speed
decreases as the particles lose mass, until the threshold for fragmentation is reached.
Equating $\Delta V$ with the fragmentation speed $V_f$ gives $\tau_{\rm stop} V_{\rm con}/r \sim (V_f/V_{\rm con})^2$.  
The fragmentation velocity depends on chemical composition, size, porosity, and other factors. Here we simply use 
1 m s$^{-1}$ and 10 m s$^{-1}$, respectively, for dust and ice aggregates \citep{blum, wada, wada2}.   
The size-dependent stopping time is related to $V_f$ and $L_{\rm con}$ via
\begin{equation}\label{eq:frag1}
    \tau_{\rm stop}(a_d) \sim {4\pi V_f^2 r^3 \rho_g\over L_{\rm con}}.
\end{equation}

When the direct gravitational acceleration dominates turbulent stirring, the particle drift speed  
is instead $\Delta V = \tau_{\rm stop} GM(r)/r^2$;  hence at the threshold for fragmentation,
\begin{equation}\label{eq:frag2}
\tau_{\rm stop}(a_d) \sim {V_f r^2\over GM(r)}.
\end{equation}
The maximum particle radius $a_f(r)$ is obtained by taking the minimum of the right-hand sides of Equations
(\ref{eq:frag1}) and (\ref{eq:frag2}).   Typically we find that turbulent stirring dominates near the
Bondi radius; hence $a_f$ increases inward until Equation (\ref{eq:frag2}) dominates.

The convective motions transport embedded particles inward from the Bondi radius toward the core.
After an initial phase of collisional breakdown in the outermost envelope, the threshold size for 
fragmentation increases inward (see Section \ref{s:dustsize2} for examples).  Inflowing
particles reach this threshold only if two conditions are satisfied:  first, the mean free time for
collisions between similarly sized particles $\tau_{\rm col}$ must remain smaller than the convective time;  
and, second, the particles must not be so large and compact that that they bounce rather than stick. 
The second condition is a complicated one, in that it depends on the porosity of the particles 
\citep{blum, porosity}.  We therefore only apply the first constraint,
\begin{equation}
\tau_{\rm col} = {1\over n(a_d) 4\pi a_d^2 V_t(a_d)} < {r\over V_{\rm con}},
\end{equation}
where $n(a_d) = 3X_d\rho_g/4\rho_s a_d^3$ is the particle number density and we approximate
$X_d \equiv \bar\rho_d/\rho_g$.  In the case where convective stirring dominates, one requires
\begin{equation}\label{eq:col1}
{a_d\over a_d + 4\ell_g/9} < 4X_d^2{\rho_g\over \rho_s} {\cal M}_{\rm con} {r\over\ell_g}.
\end{equation}
Alternatively, when the particle drift is dominated by the central gravity, then one requires
\begin{equation}\label{eq:col2}
{\tau_{\rm col} V_{\rm con}\over r} \sim {3\over 4X_d} {c_g^2 r\over GM(r)} {{\cal M}_{\rm con}\over 1 + 9a_d/4\ell_g} < 1
\end{equation}
for the particles to reach the local fragmentation threshold.

To determine the local value of the particle size, we first determine $a_f$ from Equations (\ref{eq:frag1})
and (\ref{eq:frag2}).  We set $a_d = a_f$ if the corresponding collision time is short enough for the local fragmentation bound to be reached, as determined by either Equation (\ref{eq:col1}) or (\ref{eq:col2}).
Otherwise, if collisional equilibrium is lost at some radius, then we freeze the particle size interior to this radius. 
This overall method is analogous to the model that \cite{til} developed for PPDs.

\subsection{ Suppression of Silicate Rain}\label{s:rain}
\label{rain}

Our default model sets aside the effects of silicate sublimation in the inner gasesous envelope.
This represents a conservative choice of core-envelope boundary condition for embedded grains.
In Section \ref{s:evol}, we consider the alternative situation where rain-out of silicate 
particles is suppressed by re-evaporation.  This is motivated by the following
argument that these particles (of partially solid and liquid composition) will efficiently
sublimate during their descent to the core, 
as a result of the large partial pressure of silicate vapor and large atmospheric scale height.  

The silicate saturation vapor pressure increases rapidly inward, so that if the core mass is high enough, a falling rain particle must encounter an inner layer of the envelope where the vapor pressure is well below saturation.   Within
this layer, particles whose size is limited by fragmentation ($a \sim 0.1$ mm) are in the Stokes drag regime.
Then we must take into account the diffusion of silicate vapor molecules away from a sublimating particle, and
Equation (\ref{eq:pebdes1}) must be replaced by 
\citep{pruppacher79}
\begin{equation}
    {1\over a}{da\over dt} = - f_v {D_{\rm sil} \mu_{\rm sil}
    \over \rho_s a^2 \kB T} (P_{\rm sil}^{\rm sat} - P_{\rm sil}).
\end{equation}
Here $f_v>1$ is the venting factor and the diffusion coefficient of silicate molecules through the lighter gas is \citep{lp1981} 
\begin{equation}
D_{\rm sil} \simeq 0.6 {(kT/\mu_{\rm H})^{1/2}\over 
(a_{\rm H_2} + a_{\rm sil})^2 n_{\rm H}}.
\end{equation}
In this expression, $n_{\rm H} \equiv X_{\rm H}\rho_g/\mu_{\rm H}$ is the number  density of ${\rm H}_2$ and
He molecules and $a_{\rm H_2} \sim a_{\rm sil} \sim 10^{-8}$ cm are the effective collisional radii of ${\rm H}_2$, 
SiO and ${\rm O}_2$.  Then one finds
\begin{equation}
    {1\over a}{da\over dt} \simeq -3\times 10^{-2} f_v
    \left({a\over {\rm mm}}\right)^{-2}
    \left({T\over 2000~{\rm K}}\right)^{1/2}
    {X_{\rm sil}\over X_{\rm H}} {P_{\rm sil}^{\rm sat}
     - P_{\rm sil}\over P_{\rm sil}}\quad {\rm s^{-1}}.
\end{equation}
The particle would take $\sim 10^6$ seconds to fall at the fragmentation speed ($\sim 1$ m/s) through the inner $10^3$ 
km of the envelope, meaning that it must sublimate well before reaching the core-envelope boundary.   The relative 
efficiency of sublimation in this context, compared with the incidence of `dry rain' in the Earth's atmosphere, can be 
ascribed to a combination of two factors:  the relatively high silicate vapor mass fraction (typically 
$X_{\rm sil} \sim (0.3-1)X_{\rm H}$ when refluxing of pebble debris from the envelope begins to balance pebble accretion) and
an infall time time a factor $\sim 10^3$ times larger than for raindrops.  Of course, much larger planetesimals that 
survive transport through the envelope may still inject solid/liquid silicates into the core.

\section{Envelope Expansion in Response to \\ Planetesimal Heating and Dust Loading:  Semi-Analytic Treatment}\label{s:expand}

As a prelude to the numerical investigation described in Sections \ref{s:model} and \ref{s:results}, we 
now present a semi-analytic model of an envelope with a low dust density and outer radiative structure.  This model
connects smoothly to a uniform medium of a fixed temperature $T_{\rm disk}$ and gas density $\rho_{g,\rm disk}$.
As the dust abundance rises, the inner convective layer grows in size, and eventually overwhelms the
Bondi radius.  Assuming that the dust abundance in the outer radiative layer is tied directly to the solid
accretion rate, we can work out the critical value of $\dot M_p$ above which convection extends beyond the
Bondi radius, some refluxing of solids will occur, and the self-consistency of the model breaks down.
We next quantify the reduction in dust size due to convective stirring, and demonstrate how a convective
layer that traps accreted solids may expand in response to deep heating by large planesimals.

\subsection{Two-Layer Convective-Radiative Envelope}\label{s:2layer}

Even a small amount of embedded dust acts as a powerful coolant, in the sense that any excess of 
gas temperature over ambient radiation temperature is rapidly converted to thermal radiation via 
collisions between gas molecules and grains (e.g., 
\citealt{cg1997}).  The presence of an ambient blackbody radiation field 
of temperature $T_{\rm disk}$ is therefore assumed, with an internal energy amounting to a small fraction
$\sim 10^{-5}T_{\rm disk,2}^3\rho_{g,\rm disk,-11}^{-1}$ of the ambient gas energy density.
The temperature is buffered near $T_{\rm disk}$ in the outer bound envelope, around the Bondi radius, 
and grows in the inner, optically thick envelope.

In a state of spherical hydrostatic and radiaive equilibrium at local temperature $T_g$ and 
uniform luminosity $L_{\rm rad}$,
\begin{eqnarray}\label{eq:hyd1}
{dP_g\over dr} &=& {d\over dr}\left({\rho_gkT_g\over\mu_g}\right) = -GM_c{\rho\over r^2}; 
\quad\quad \rho_g = (1-X_d)\rho;\nn
{L_{\rm rad}\over 4\pi r^2} &=& -{4\over 3\kappa_d(T_g)X_d\rho}{d(\sigma_{\rm SB}T_g^4)\over dr}.
\end{eqnarray}
The opacity $\kappa_d(T) \propto T^{1+\beta}$ is per unit mass of dust.
We first consider the case of uniform dust mass fraction $X_d$ (although the equations presented here
  do not depend on this assumption).
We rescale radius to $R_{\rm B}$, temperature to $T_{\rm disk}$, and gas density to 
$\rho_{g,\rm disk}$.  Then the preceding equations can be written
\begin{eqnarray}\label{eq:dimstruc}
{d(\widetilde\rho_g\widetilde T_g)\over d\widetilde r} &=& -{\widetilde\rho_g\over(1-X_d)\widetilde r^2};\nn
{\widetilde T_g^{2-\beta}\over\widetilde\rho_g} {d\widetilde T_g\over d\widetilde r} &=& - 
{\varepsilon_{\rm rad}\over \widetilde r^2},
\end{eqnarray}
where
\begin{equation}
\varepsilon_{\rm rad} = {3\tau_{\rm B}\over 64\pi} {L_{\rm rad} c_{g,\rm disk}^4\over (GM_c)^2 \sigma_{\rm SB}
T_{\rm disk}^4} = {3\tau_{\rm B}\over 64\pi} {c_{g,\rm disk}^4\over GR_c t_{{\rm acc},c} \, \sigma_{\rm SB}
T_{\rm disk}^4}
\end{equation}
and the luminosity is expressed in terms of a core accretion time,
\begin{equation}
L_{\rm rad} = {GM_c^2\over R_c t_{{\rm acc},c}}.
\end{equation}
Taking $\kappa_d = 230 (T/100~{\rm K})^{1+\beta}$ cm$^2$ g$^{-1}$ (see Appendix \ref{s:kapgr}),
the characterisic optical depth at the Bondi radius is
\begin{equation}
\tau_{\rm B} \equiv {X_d\over 1-X_d} \kappa_d(T_{\rm disk}) \rho_{g,\rm disk} {GM_c\over c_{g,\rm disk}^2}
  = 0.77\,\left({X_d\over 1-X_d}\right)_{-2} \rho_{g,\rm disk,-11} T_{\rm disk,2}^\beta \left({M_c\over 0.3~M_\oplus}\right)^{2/3}
\left({t_{{\rm acc},c}\over {\rm Myr}}\right)^{-1},
\end{equation}
and 
\begin{equation}\label{eq:eprad}
\varepsilon_{\rm rad} = 0.024\,\left({X_d\over 1-X_d}\right)_{-2} {\rho_{g,\rm disk,-11}\over T_{\rm disk,2}^{2-\beta}}
\left({M_c\over 0.3~M_\oplus}\right)^{2/3}\left({t_{{\rm acc},c}\over {\rm Myr}}\right)^{-1}.
\end{equation}

\begin{figure}
\epsscale{1.1}
\plottwo{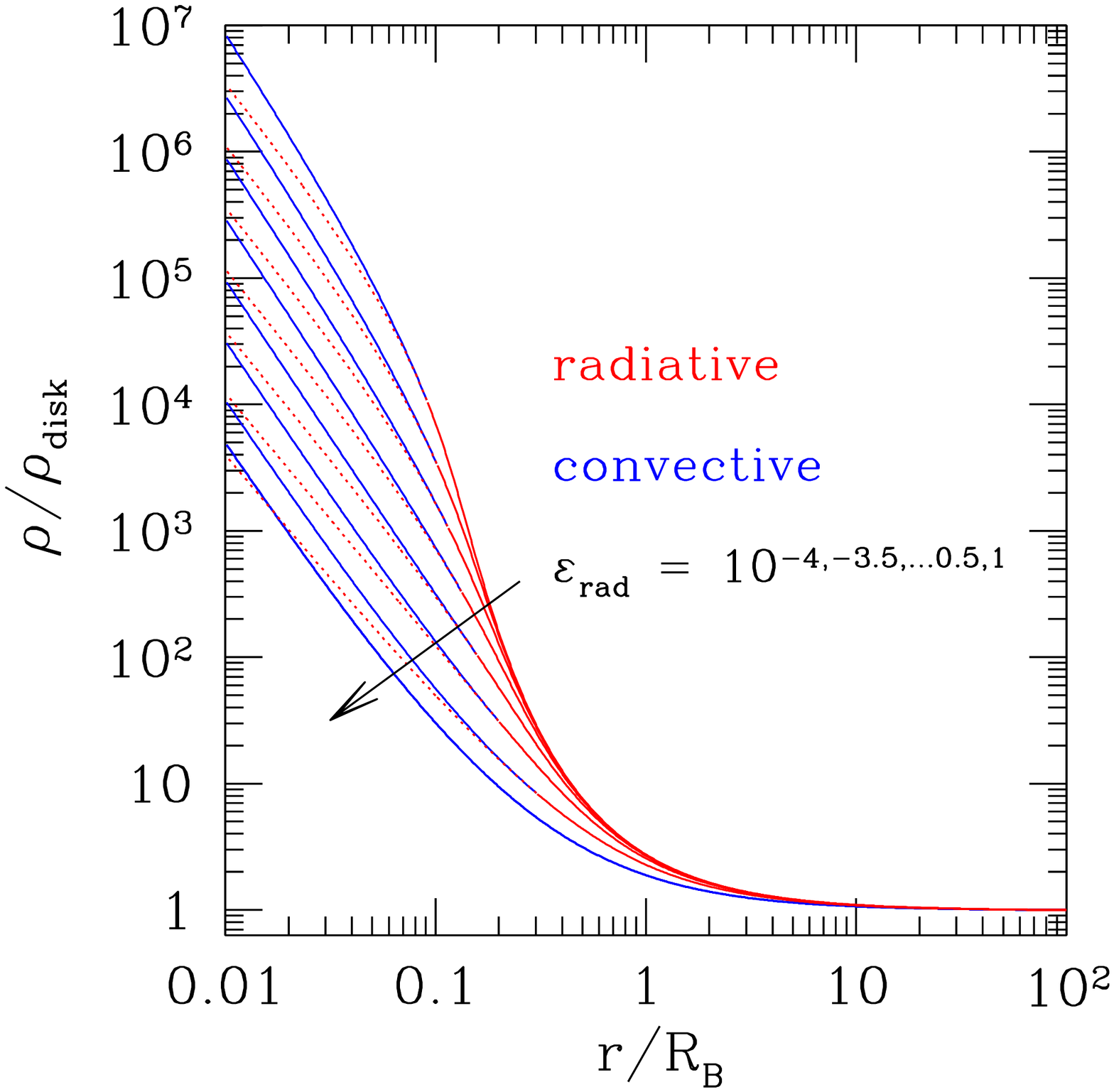}{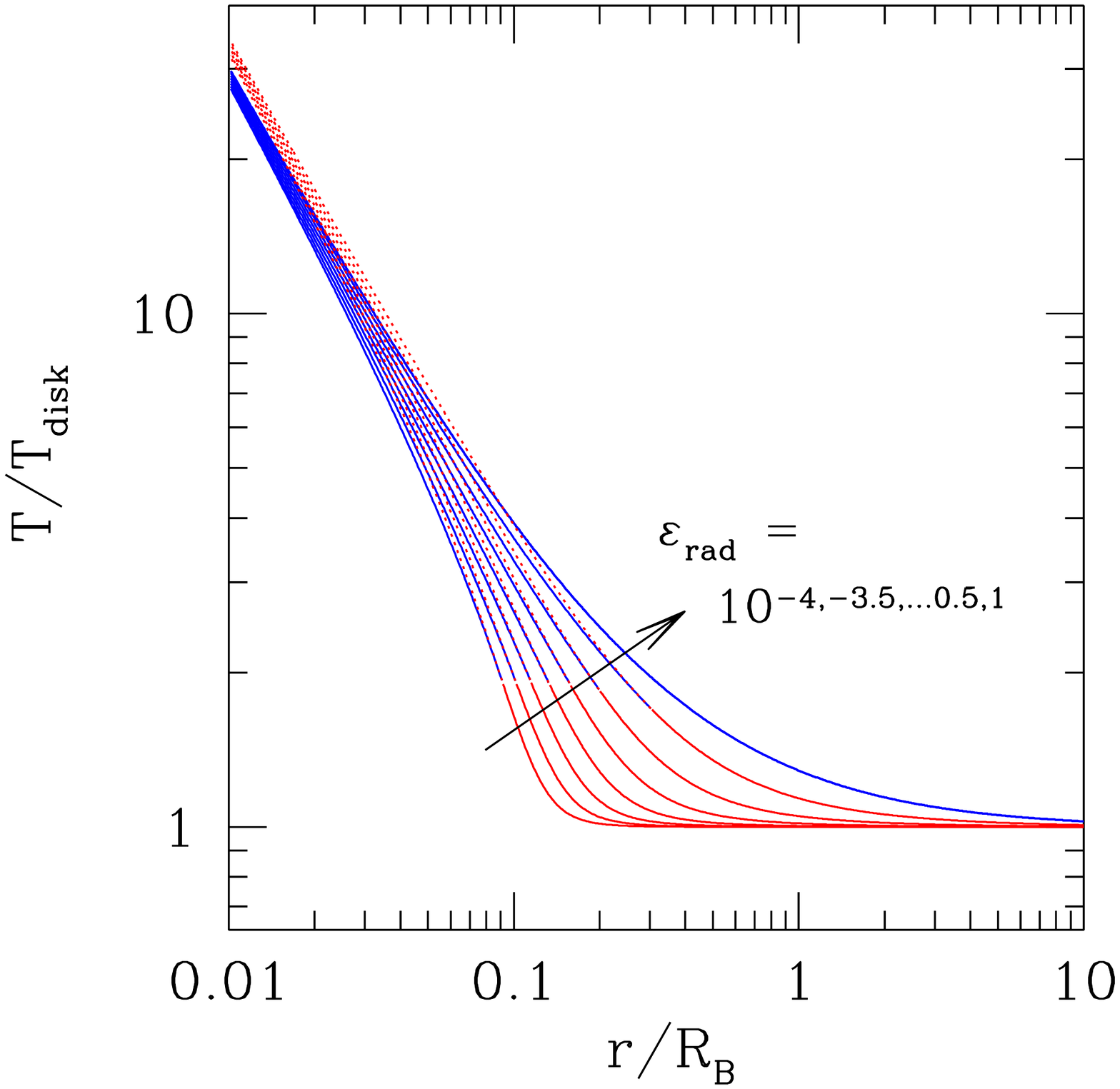}
   \caption{Structure of a dusty, hydrostatic envelope accreted from a uniform gaseous medium onto a point mass.
A uniform radiative luminosity passes through the envelope from the core boundary. The envelope structure, determined
by Equations (\ref{eq:hyd1}) and (\ref{eq:dimstruc}), is parameterized by the quantity $\varepsilon_{\rm rad}$
(Equation (\ref{eq:eprad})) which is proportional to the dust mass fraction $X_d$ and the radiative luminosity.
When this parameter is small, the outer envelope is radiative and the inner envelope
is much denser than a fully adiabatic model, by a factor $\sim \varepsilon_{\rm rad}^{-1}$.  The inner envelope
is convective if the core is sufficiently dense or the ambient medium cool enough.  The radiative-convective
boundary moves outward as the dust loading and/or the radiative flux increases, until a critical point is
reached (Equation (\ref{eq:xdmin})) where the entire bound envelope convects.  The models with $\varepsilon_{\rm rad}
= 10^{-0.5, 0, 0.5, 1}$ are fully convective.
The dotted red curves show the inward extrapolation of the radiative solution.}
\end{figure}\label{fig:radprof}

The solution to Equations (\ref{eq:dimstruc}) is shown in Figure \ref{fig:radprof} for $\beta = 0$ and
a range of values of the parameter $\varepsilon_{\rm rad}$ ($10^{-4}$, $10^{-3.5}$, $10^{-3}$, ... $10^{0.5}$, 10).
When $\varepsilon_{\rm rad} \ll 1$, the temperature remains flat some distance inside $R_{\rm B}$,
whereas the density exponentiates inward,
\begin{equation}
\widetilde\rho_g(r) = {\rho_g(r)\over \rho_{g,\rm disk}} \simeq \exp\left[{R_{\rm B}\over (1-X_d)\,r}\right].
\end{equation}
The entropy grows with radius in this outer zone (red solid curves in Figure \ref{fig:radprof}).
The outer envelope forms a stable radiative layer, and represents a ``buoyancy barrier'' such as
is seen in the simulations of \cite{kurokawa} (which however did not include planetesimal heating or dust
production by pebble fragmentation).  The density in the inner power-law core of the envelope 
is enhanced by a factor $\varepsilon_{\rm rad}^{-1}$.

The inner power-law profile obtained from Equations (\ref{eq:dimstruc}), which is shown as the dotted red lines in 
Figure \ref{fig:radprof}, is convectively unstable.  
The adiabatic extension of the outer radiative profiles appears as the blue curves in Figure \ref{fig:radprof}.
Convection sets in when
$d\ln T/dr < (\gamma-1)d\ln\rho/dr$, corresponding to $(\gamma-1)(2-\beta) < 1$
and $\beta > -{1\over 2}$ for an adiabatic index $\gamma \simeq 1.4$.  Then the inner density profiles
scales as $\rho(r) \propto r^{-1/(\gamma-1)}$.  The inner and outer solutions cross at 
\begin{equation}
R_{\rm rad-con} \simeq
R_{\rm B} \left\{(1-X_d)\ln\left[{1\over \varepsilon_{\rm rad}[(3-\beta)(1-X_d)]^{3-\beta}}\right]\right\}^{-1}.
\end{equation}

A general feature of this simplified solution, which survives in the more complete numerical treatment,
is that a rising dust abundance increases $\varepsilon_{\rm rad}$, reduces the density in the inner envelope,
and pushes the radiative-convective boundary $R_{\rm rad-con}$ outward.  One observes 
in Figure \ref{fig:radprof} that
the profile is entirely convective for $\varepsilon_{\rm rad} \gtrsim 0.3$.  

In particular, to determine whether the profile is
convective at the Bondi radius, we substitute $\widetilde\rho_g$, $\widetilde T
\simeq 1$ into Equations (\ref{eq:dimstruc}) and find the condition
\begin{equation}
\varepsilon_{\rm rad} > {\gamma-1\over\gamma(1-X_d)}\quad\quad({\rm fully~convective}).
\end{equation}
Since $\varepsilon_{\rm rad}$ is proportional to $X_d/(1-X_d)$, 
e.g. $\varepsilon_{\rm rad} = \varepsilon_{\rm rad}^* X_d/(1-X_d)$, this condition is equivalent to 
\begin{equation}\label{eq:xdmin}
X_d > {\gamma-1\over\gamma\varepsilon_{\rm rad}^*} = 0.12\,{T_{\rm B,2}^2\over\rho_{g,\rm B,-11}}
 \left({t_{{\rm acc},c}\over {\rm Myr}}\right)\left({M_c\over 0.3~M_\oplus}\right)^{-2/3}.
\end{equation}
Equivalently, for an ambient temperature of $\sim 100$ K and density $\sim 10^{-11}$ g cm$^{-3}$, convective
refluxing across the Bondi radius can occur if the accretion rate rises above a modest
threshold corresponding to $t_{{\rm acc},c} \sim 10$ Myr.  This is only a few percent of the accretion rate
that is needed to build the core up to $\sim 10\,M_\oplus$ over a Myr.

\begin{figure}
\epsscale{0.7}
\plotone{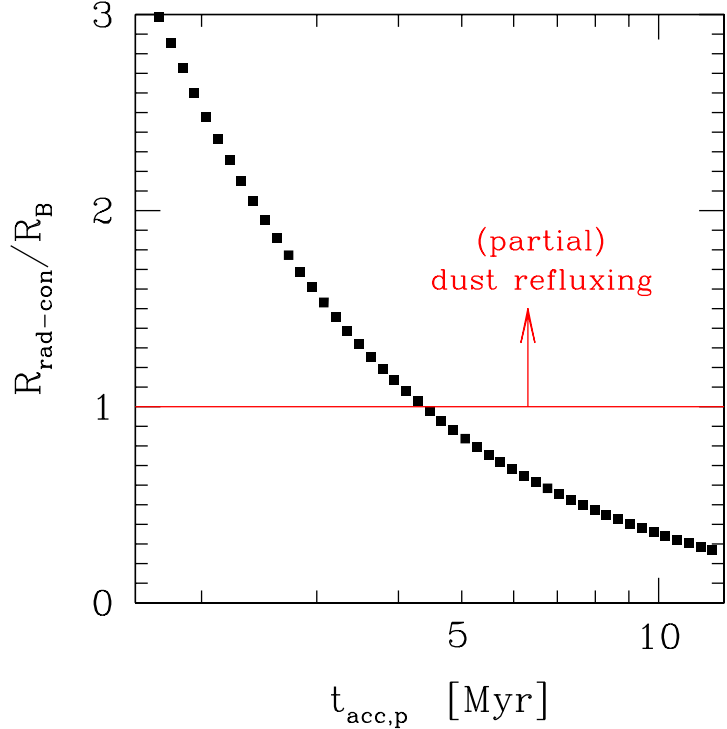}
\vskip -0.5in
   \caption{The envelope models shown in Figure \ref{fig:radprof} have been generalized to include
   the dust opacity model of Appendix \ref{s:kapgr} and a steady accretion rate of pebble debris
    $\dot M_p$.  We plot the boundary radius $R_{\rm rad-con}$ between the
    inner convection zone and the outer radiative layer, which is represented in Figure \ref{fig:radprof}
    by the transition from blue to red curves, as a function of the core e-folding time $t_{{\rm acc},p} = M_c/\dot M_p$. 
    Ambient disk temperature $T_{\rm disk} = 10^2$ K, gas density $\rho_{g,\rm disk} = 10^{-11}$ g cm$^{-3}$.  
    The core luminosity can self-consistently be approximated as $GM_c\dot M_p/R_c$ as long as the
    convective layer remains confined within the gravitationally bound envelope ($R_{\rm rad-con} \lesssim R_{\rm B}$).
    When $t_{{\rm acc},p}$ is smaller than $\sim 4$ Myr, convection extends beyond the Bondi radius, which is
    an approximate criterion for the expulsion of some pebble debris.  Otherwise all the trapped pebble material 
    should be accreted.}
\end{figure}\label{fig:radprof2}

This model can be easily extended to allow for a steady, inward radial drift of solid dust particles through 
the outer radiative layer of the envelope.   The fragmentation of pebbles in the outer envelope sources
dust grains, which must drift inward in the absence of convective motions. A simple first estimate 
of the dust density is obtained by balancing the dust mass flux with the imposed pebble mass flux,
and assuming that the dust speed is equal to the fragmentation speed,
\begin{equation}
\dot M_p = 4\pi r^2 V_f \cdot n_d {4\pi\over 3} a_d^3 \rho_s.
\end{equation}
As a result, the dust size is typically $a_d < \lambda_{\rm max}(T)$ (Equation (\ref{eq:admax})), 
and so the dust opacity is independent of $a_d$ (see Appendix \ref{s:kapgr})
and depends mainly on $\dot M_p$ and $V_f$.  The dust mass fraction solves
\begin{equation}
    {X_d\over 1-X_d} = {13.6\over \widetilde\rho_g  \widetilde r^2}  {T_{\rm disk,2}^2\over 
    \rho_{g,\rm disk,-11} V_{f,2}}  \left({\dot M_p\over 10^{-6}\,M_\oplus~{\rm yr}^{-1}}\right)
    \left({M_c\over 0.3~M_\oplus}\right)^{-2}.
\end{equation}

Figure \ref{fig:radprof2} shows how the boundary $R_{\rm rad-con}$ between the inner convection
zone and the outer radiative layer depends on the imposed solid accretion rate, for
$M_c = 0.3\,M_\oplus$, $T_{\rm disk} = 10^2$ K, and $\rho_{g,\rm disk} = 10^{-11}$ g cm$^{-3}$.
The core luminosity is taken to be $GM_c\dot M_p/R_c$, which is self-consistent as long as the
convection zone remains confined approximately to the Bondi sphere.   One observes that refluxing
sets in when the core e-folding time $t_{{\rm acc},p} \lesssim 4$ Myr; otherwise all the trapped
pebble material should be accreted.  This constraint is in good agreement with the time-dependent envelope
calculations presented in Section \ref{s:results}.  By contrast, the envelope models of
  \cite{ormel} and \cite{mordaop} find more compact convective zones because they allow for adhesive
  but not fragmenting collisions between grains, and therefore incorporate a much larger grain size
  and inward drift speed.

\subsection{Dust Size Reduction by Turbulent Stirring and Secular Drift}\label{s:dustsize2}

To proceed further, one needs a model for the evolution of the dust size, which is described
in Section \ref{s:dustsize} and implemented in the numerical model of Sections \ref{s:model} and \ref{s:results}.
Here we estimate analytically the equilibrium dust size in a fully developed convective zone, both in the outer
parts where dust particles are in the Epstein drag regime ($a_d < \ell_g$) and the turbulent acceleration
dominates, and in the inner envelope where the Stokes drag regime.  Our estimates agree well with the
model results. 

In a fully convective state, the spherical temperature and density profiles around the core take the
simple form
\begin{equation}
    \begin{split}
    T(r) &= T_{\rm disk}\left[1+{(\gamma-1)R_{\rm B}\over \gamma r}\right] \\
    \rho_g(r) & = \rho_{g,\rm disk}\left[1+{(\gamma-1)R_{\rm B}\over \gamma r}\right]^{1/(\gamma-1)}.
    \end{split}
\end{equation}
These profiles connect smoothly to a uniform ambient medium.
In this section, we choose a ratio of specific heats $\gamma = 1.4$ and mean molecular weight
$\mu = 2.3\,m_u$, as appropriate for an atmosphere dominated by H$_2$/He.

The luminosity is normalized in terms of the core accretion time $t_{{\rm acc},c}$,
\begin{equation}\label{eq:lcore}
L_{\rm con} = 4\pi r^2 \rho_g V_{\rm con}^3 \equiv GM_c^2/R_ct_{{\rm acc},c}.
\end{equation}
(This expression assumes high convective efficiency, the justification for which is given in
Appendix \ref{s:coneff}.)   The convection remains subsonic everywhere in the atmosphere as 
long as $L_{\rm core}$ ($t_{{\rm acc},c}$) remain below (above) a critical value:
\begin{equation}
{\cal M}_{\rm con}^3 \equiv \left({V_{\rm con}\over c_g}\right)^3 = 
{49\over 16\pi}  {c_{g,\rm disk}\over G\rho_{g,\rm disk} R_c t_{{\rm acc},c}}
   {(7r/2R_{\rm B})^2\over (1+7r/2R_{\rm B})^4}.
\end{equation}
This is maximized at $r = 2R_{\rm B}/7$, where ${\cal M}_{\rm con} < 1$ as long as
\begin{equation}
    t_{{\rm acc},c} > 1.7\times 10^5\, T_{\rm disk,2}^{1/2} \rho_{g,\rm disk,-11}^{-1} R_{c,9}^{-1}\quad{\rm yr}.
\end{equation}
and
\begin{equation}
    L_{\rm con} < 4.1\times 10^{25}\,\left({M_c\over 0.3~M_\oplus}\right)^2 T_{\rm disk,2}^{-1/2}\rho_{g,\rm disk,-11}\quad{\rm erg~s^{-1}.}
\end{equation}

To determine the marginally fragmenting grain size in the outer, turbulent convective envelope, we
make use of Equation (\ref{eq:frag1}).  Substituting for $L_{\rm con}$ in terms of the accretion time 
(Equation (\ref{eq:lcore})) gives
\begin{eqnarray}\label{eq:adust}
a_d &\sim& 1.1\times 10^{-7}\,V_{f,2}^2\,\rho_{s,0}^{-1}
\,T_{\rm disk,2}^{-5/2}\,\rho_{g,\rm disk,-11}^2
\left({M_c\over 0.3~M_\oplus}\right)^{4/3}
\left({t_{{\rm acc},c}\over {\rm Myr}}\right) \nn 
&& \quad \times \left({7r\over 2R_{\rm B}}\right)^{-5/2}
\left(1+{7r\over 2R_{\rm B}}\right)^{11/2}\quad{\rm cm}.
\end{eqnarray}
At $r \sim 2R_{\rm B}/7$ the coefficient works out to $\sim 0.03$ $\mu$m for silicate grains with fragmentation speed
$\sim 1$ m s$^{-1}$, increasing by a factor $\sim 10^2$ for icy grains with fragmentation speed
10 times higher. 

In the inner envelope, the dominant acceleration acting on grains can either be due to turbulence 
or the central core gravity, with the latter dominating when the planetesimal accretion rate is
relatively low ($\lesssim 3\times 10^{-8}\,M_\oplus$ yr$^{-1}$).  In the latter case, taking
$V_f = 1$ m s$^{-1}$ as appropriate for the silicate grains that would be present at temperatures
well above 100 K, one has from Equations (\ref{eq:ts}) and (\ref{eq:frag2})
\begin{equation}
    a_d \sim \left({\mu_g r^2 c_g V_f \over GM_c \rho_s\sigma_{\rm H}}\right)^{1/2}
      = 5\times 10^{-3} V_{f,2}^{1/2}\left({r\over R_c}\right)^{3/4} \quad {\rm cm}\quad \quad ({\rm Stokes}).
\end{equation}
This shifts to a scaling $a_d(r) \propto r$ when the grain size drops below $\ell_g \propto r^{5/2}$
(Epstein regime).

\begin{figure}
\epsscale{0.9}
\plotone{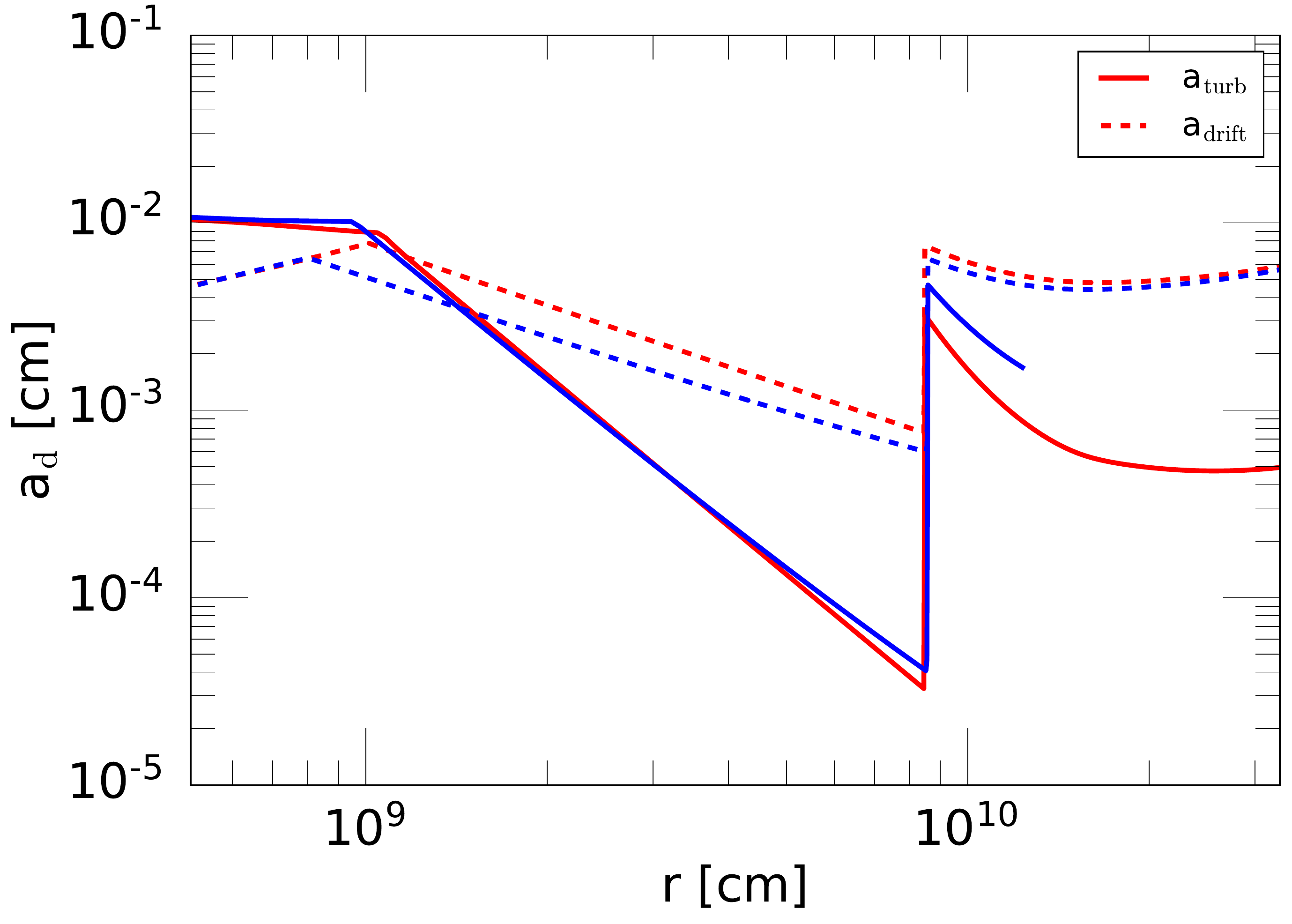}
   \caption{Dust grain radius $a_{\rm d}$ as 
   limited by turbulent fragmentation (Equation (\ref{eq:frag1}); solid curves) or by radial drift 
   (Equation (\ref{eq:frag2}); dashed curves), at $t=10^3$ yr in our default model with $\dot M_p = 10^{-6}\,M_\oplus$ yr$^{-1}$
and $\dot M_{\rm plan} = 0$ (red curves) or $\dot M_p = 10^{-7}\,M_\oplus$ yr$^{-1}$
(blue curves).  The upward shift in $a_d$ at $\sim 10^{10}$ cm represents the appearance of water ice in the grains;
we take $V_f = 1$ m s$^{-1}$ for silicate grains and 10 m s$^{-1}$ for icy grains.  
The equilibrium grain size (Figure \ref{fig:dustsize}) is given by the minimum of the solid and
dashed curves.
The cutoff in the solid blue line represents the convective-radiative boundary.}
    \label{fig:psizer}
\end{figure}

\begin{figure}
\epsscale{0.9}
\plotone{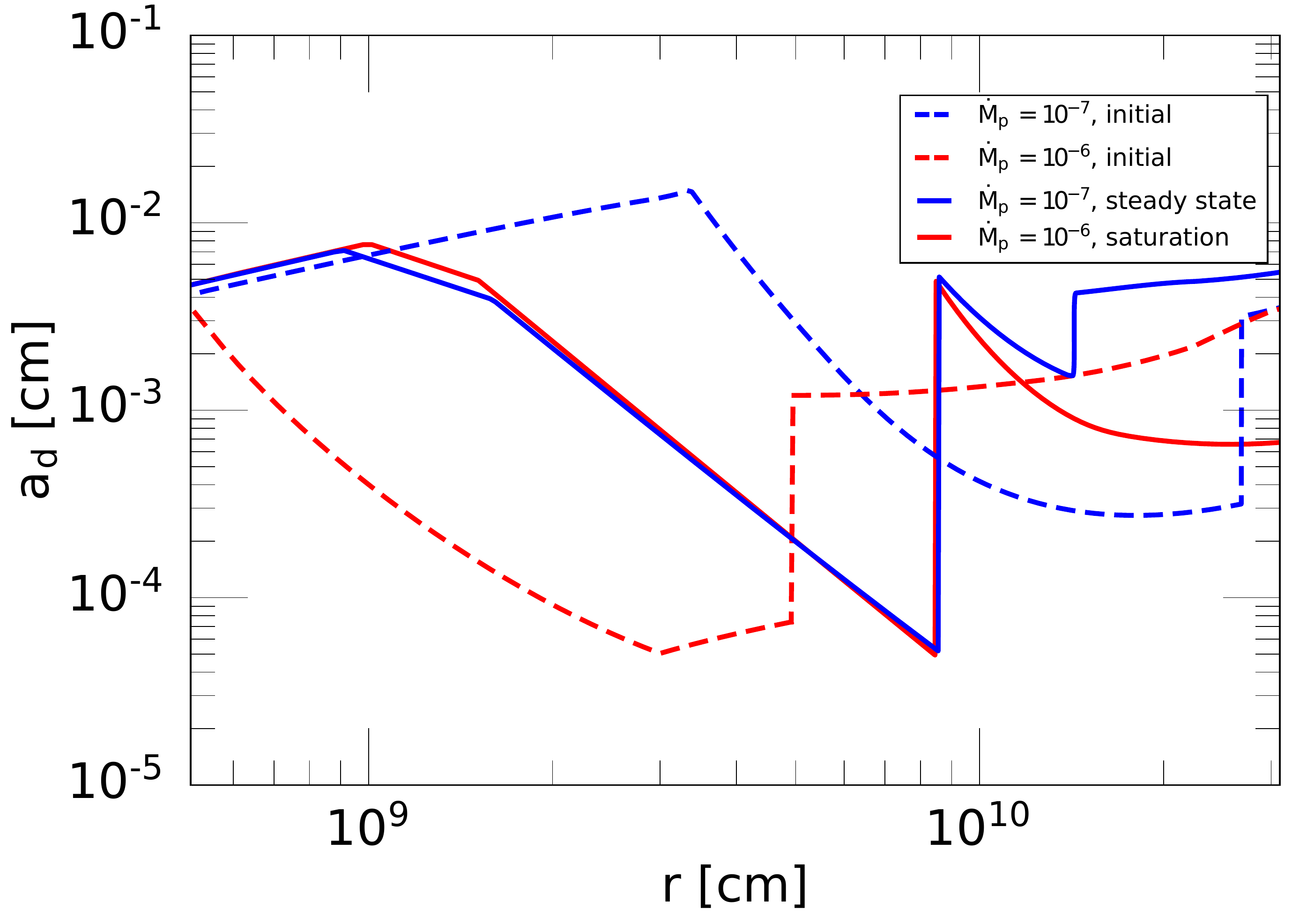}
\caption{Evolution of the dust radius $a_d$ between $t = 50$ yr (``initial'') and $10^3$ yr (``steady state'')
 in our time-dependent envelope models.
The dust size prescription allows for both collisionally-induced fragmentation and 
  sticking, with $a_d$ typically attracted to the marginally-fragmenting state.
Comparison with Figure \ref{fig:psizer} shows that, at steady state (solid curves), dust size is limited by drift-induced fragmentation in the 
inner envelope for both $\dot M_p$ values, while convective turbulent fragmentation dominates in the outer envelope. 
However at $t = 50$ yr (dashed curves), the low-$\dot M_p$ model maintains weak turbulent stirring, and so drift-dominated drag
extends farther out before shifting to turbulent stirring.  In the high-$\dot M_p$ model, 
the density remains low enough in the inner envelope that the dust is in the Epstein drag regime and is convectively fragmented.}
\label{fig:dustsize}
\end{figure}

Figures \ref{fig:psizer} and \ref{fig:dustsize} give examples of
how the fragmentation model of Section \ref{s:dustsize}
is implemented in our numerical model.  The solid curves in Figure \ref{fig:psizer} show the size as limited by fragmentation in
response to turbulent stirring, and the dashed curves in response to radial drift in the gravitational
field.  The large break at $\sim 10^{10}$ cm represents a transition from amorphous silicate-rich 
conglomerates with fragmentation speed $V_f \sim 1$ m $s^{-1}$ to ice-rich grains which stick at higher
speeds, $V_f \sim 10$ m s$^{-1}$.  The innermost breaks in both red and blue curves represent a transition
from an inner Stokes drag regime to the outer Epstein drag regime.  

\subsection{Expansion of a Dust-Loaded Convective Layer Driven by Deep Accretion Heating}

Starting from an initial envelope profile with an outer radiative layer, our numerical simulations
show that most of the pebble mass is deposited in this outer layer.  The differential 
growth in mean molecular weight $\mu$ in the radiative layer, relative to the interior convective zone,
creates an additional source of instability near the radiative-convective boundary.  This enhances the
growth of the convective layer both by reducing the radiative energy flux and through salt-finger mixing
across the boundary.  Note also that if a majority of the pebble material were deposited in the deep 
convective layer, it would rapidly mix within a much larger mass as compared with the outer radiative shell,
thereby limiting the relative growth of $\mu$ in the convective layer.  

Such salt-fingering effects are not taken into account in our calculations. 
We use a simplified transport model in which the temperature gradient is the minimum of the adiabatic gradient
and the radiative gradient (Section \ref{s:struc}).

The opposing situation, where the mean molecular weight grows fastest in the convective layer, 
leading to the appearance of a stabilizing mean molecular weight gradient at its outer boundary,
is typically not realized in our simulations.  Even in such a circumstance, the rapid rise in
dust opacity in the convective layer, combined with a small continuing injection of heat,
would force the convective layer to expand.  Supposing a fixed mass $\Delta M$ of lower mean 
molecular weight material to sit outside a radius 
$R_{\rm con}$, a fractional change $\delta s/s$ in the entropy function $s = P/\rho^\gamma$ drives
a relatively large expansion $\delta R_{\rm con}/R_{\rm con} \sim 5\delta s/s$, where $M_{\rm con}$ is the 
mass of the convective layer. 

To show this, consider a situation where
the convective layer is sandwiched between a dense core (radius $R_c$), and thin 
outer radiative layer positioned at $r > R_{\rm con} \gg R_c$.  The total gravitating mass is dominated by 
the core mass 
$M_c$, and the masses $M_{\rm con}$ and $M_{\rm rad}$ of the convective and radiative layers are taken to be fixed.
The convective layer is adiabatic, with a pressure profile $P(r) = P(R_{\rm con})[\rho/\rho(R_{\rm con})]^\gamma$.  Substituting
this equation of state into the equation of hydrostatic equilibrium, and defining $s = P/\rho^\gamma$, one finds
\begin{equation}
    \rho(r) = \rho(R_{\rm con}) + \left[{\gamma-1\over \gamma}{GM_c\over s}
    \left({1\over r}-{1\over R_{\rm con}}\right)\right]^{1/(\gamma-1)}.
\end{equation}
The convective mass is
\begin{equation}\label{eq:mcon}
    M_{\rm con} = \int_{R_c}^{R_{\rm con}} 4\pi r^2 \rho(r) dr \simeq
         {4\pi\over 3} R_{\rm con}^3 \left[\rho(R_{\rm con}) + 
             3 f(\gamma) \left({\gamma-1\over\gamma}{GM_c\over s R_{\rm con}}\right)^{1/(\gamma-1)}\right],
\end{equation}
where $f(\gamma) \equiv \int_0^1 x^2 (x^{-1}-1)^{1/(\gamma-1)} dx$.   Applying pressure balance across the outer convective-radiative
boundary gives $ s \cdot [\rho(R_{\rm con})]^\gamma = GM_c M_{\rm rad}/4\pi R_{\rm con}^4 $.  
One sees that
both terms in Equation (\ref{eq:mcon}) are functions of $s\cdot R_{\rm con}^{4-3\gamma}$. 
Taking $\delta M_{\rm con} = 0$ then gives
\begin{equation}
 {\delta R_{\rm con}\over R_{\rm con}} = {1\over 3\gamma-4}{\delta s\over s} = {1\over 3\gamma-4}\delta[\ln s].
\end{equation}
This implies a strong dependence of $R_{\rm con}$ on $s$, namely $\delta R_{\rm con}/R_{\rm con} \sim 5\delta s/s$,
for an equation of state dominated by diatomic gases.

For all of the preceding reasons, we neglect the influence of a mean molecular weight 
gradient on the radiative-convective transition (explored  for example by
\citealt{theado, leconte, vazan2016,helled}), except through its indirect effect on the opacity.

\section{Numerical Model}
\label{s:model}

We now introduce a spherical hydrostatic model of the gaseous envelope surrounding a growing planetary core, 
which is embedded in a PPD.  A spherical approximation to the envelope structure is well
motivated inside the core Bondi radius.  The envelope 
metallicity generally exceeds the Solar value, being dominated in the outer parts by solid grains and in the inner parts
by water and silicate vapor derived by pebble accretion.  A luminosity $L_{\rm tot}(r)$ is transported through the
envelope by a combination of convection and radiative diffusion.

\subsection{Atmospheric Structure}\label{s:struc}

The evolution of the envelope is calculated using a sequence of static models,
governed by the equations of hydrostatic equilibrium and thermal diffusion,
\begin{equation}\label{eq:dpdr}
{dP\over dr} = -GM(r){\rho\over r^2} = -GM(r){\rho_g + \bar\rho_d \over r^2},
\end{equation}
and
\begin{equation}\label{eq:dtdr}
\frac{dT}{dr}=\frac{dP}{dr}\frac{T}{P} \nabla.
\end{equation}
Here $P$ and $T$ are total pressure and temperature, and $M(r)$ the enclosed mass. 
The temperature gradient $\nabla \equiv {d\log T}/{d\log P}$ is taken to be the minimum of
the radiative and adiabatic gradients,
\begin{equation}\label{eq:nab}
\nabla = {\rm min}(\nabla_{\text{rad}},\nabla_{\text{ad}}),
\end{equation}
as defined below.  The gas mass density is determined from $P$ and $T$,
\begin{equation}
    \rho_g = \rho_g(P,T).
\end{equation}
The mean dust mass density $\bar\rho_d = X_d\rho_g/(1-X_d)$ is evolved separately, as described below.

\subsection{Effects of Ice and Silicate Sublimation}

We numerically split the envelope into two parts.  (i) An interior zone where water ice is absent,
where we use tabulated equations of state (EOSs).  For water vapor we use the NIST/STEAM V3.0 EOS \citep{nist},
\textcolor{black}{and for H$_2$/He the EOS of \cite{milit1,milit2}}.  (ii) The outer envelope, where ice is present but the gases
are dilute, and a simpler ideal gas approximation is adopted for both water vapor and H$_2$/He.

The density of a hydrogen/helium/water vapor mixture is calculated using the additive-volume rule (e.g.  \citealt{fontaine}),
\begin{equation}\label{eq:addvol}
{1\over\rho(P,T)} = {X_{\rm H}+X_{\rm H_2O,v}\over\rho_g(P,T)} =
{X_{\rm H}\over \rho_{\rm H}(P,T)} + {X_{\rm H_2O,v}\over \rho_{\rm H_2O,v}(P,T)}.
\end{equation}
Here $X_{\rm H}$ and $X_{\rm H_2O,v} = 1-X_{\rm H}-X_{\rm ice}-X_{\rm sil}$ are the mass fractions of H$_2$/He and 
water vapor;  $X_{\rm ice}$ and $X_{\rm sil}$ are the mass fractions in water ice and silicates.
Turbulent mixing adjusts the total H$_2$O mass fraction in vapor and ice components to a nearly
uniform value in the convection zone, $X_{\rm H_2O} = X_{\rm ice} + X_{\rm H_2O,v} \simeq$ const.

The sublimation of water ice extracts heat from the gas, according to the heat equation
\begin{equation}
T ds = C_p dT - \frac{1}{\rho}dP = l_{\rm H_2O} \ dX_{\rm ice}
\end{equation}
where $s$ is the specific entropy, $l_{\rm H_2O} = 2.3\times 10^{10}$ erg g$^{-1}$ is the
latent heat of sublimation and $C_p$ is the specific heat at constant pressure,
\begin{equation}
\label{cp}
C_p = (X_{\rm H_2O}-X_{\rm ice}){4\kB\over \mu_{\rm H_2O}} + X_{\rm H}{7\kB\over2\mu_{\rm H}}.
\end{equation}
The coefficients here correspond to 2 rotational degrees of freedom for H$_2$ and
3 for H$_2$O.
We take $\mu_{\rm H} = 2.3\,m_u$ to represent a H$_2$/He mixture, and $\mu_{\rm H_2O}$ is the mass of a water
molecule.  
Substituting this into the heat equation above and inserting the equation of hydrostatic equilibrium gives a fourth
identity to supplement Equations (\ref{eq:dpdr}), (\ref{eq:dtdr}), (\ref{eq:nab}) and (\ref{eq:addvol}),
\begin{equation}
\label{dxi}
l_{\rm H_2O} \ \frac{dX_{\rm ice}}{dr} = C_p\frac{dT}{dr} + g.
\end{equation}

The adiabatic gradient $\nabla_{\text{ad}}$ is now defined as follows.
The total pressure in the water sublimation zone is
\begin{equation}\label{eq:ptot}
P = P_{\rm H_2O}^{\rm sat}(T) + {X_{\rm H} \rho  T \over  \mu_{\rm H}} =
P_{\rm H_2O}^{\rm sat}(T)\bigg( 1 + \frac{X_{\rm H}}{X_{\rm H_2O} - X_{\rm ice}}\frac{\mu_{\rm H_2O}}{\mu_{\rm H}} \bigg),
\end{equation}
where $P_{\rm H_2O}^{\rm sat} = P_{\rm H_2O}^0e^{-l_{\rm H_2O}\mu_{\rm H_2O}/kT}$ is the saturation pressure,
$P_{\rm H_2O}^0 = 3.6\times 10^{13}$ dyne cm$^{-2}$.
Differentiating Equation (\ref{eq:ptot}) with respect to radius and combining with the equation of hydrostatic equilibrium
gives
\begin{eqnarray}
- \rho g = \frac{dP_{\rm H_2O}^{\rm sat}}{dT}\frac{dT}{dr} \bigg(1 + \frac{X_{\rm H}}{X_{\rm H_2O} - X_{\rm ice}}
\frac{\mu_{\rm H_2O}}{\mu_{\rm H}}\bigg) \nn
 + \quad P_{\rm H_2O}^{\rm sat}\,{X_{\rm H}\over (X_{\rm H_2O}-X_{\rm ice})^2}
    \frac{\mu_{\rm H_2O}}{\mu_{\rm H}}\frac{dX_{\rm ice}}{dr}.
\end{eqnarray}
Combining this with Equation (\ref{dxi}) gives
\begin{equation}\label{eq:dtdp}
\frac{dT}{dP}\biggr|_{\rm ad} = \frac{1 + {\cal P}} {P_{\rm H_2O}^{\rm sat\,'}(T)
\left(1 + \frac{X_{\rm H}}{X_{\rm H_2O} - X_{\rm ice}}\frac{\mu_{\rm H_2O}}{\mu_{\rm H}}\right)
+ {\cal P}\rho C_p},
\end{equation}
where
\begin{equation}\label{eq:calP}
{\cal P} \equiv {P_{\rm H_2O}^{\rm sat}\over \rho \ell_{\rm H_2O}}
{X_{\rm H} \over (X_{\rm H_2O}-X_{\rm ice})^2}{\mu_{\rm H_2O}\over\mu_{\rm H}}.
\end{equation}
and $C_p$ is given by Equation (\ref{cp}).  Finally,
\begin{equation}
\nabla_{\text{ad}} =  \frac{P}{T} \frac{dT}{dP}\biggr|_{\rm ad}.
\end{equation}
These equations are equivalent to those derived by \cite{ingersoll69} for a wet adiabat.  
We emphasize that Equation (\ref{eq:dtdp}) is only valid in the water ice sublimation zone, where H$_2$/He gas and water
vapor are approximated as ideal gases.  In the inner envelope, where ice has entirely vanished, $\nabla_{\text{ad}}$ is
retrieved directly from the tabuled EOSs.

\begin{figure}
\begin{centering}
        \includegraphics[scale=0.75]{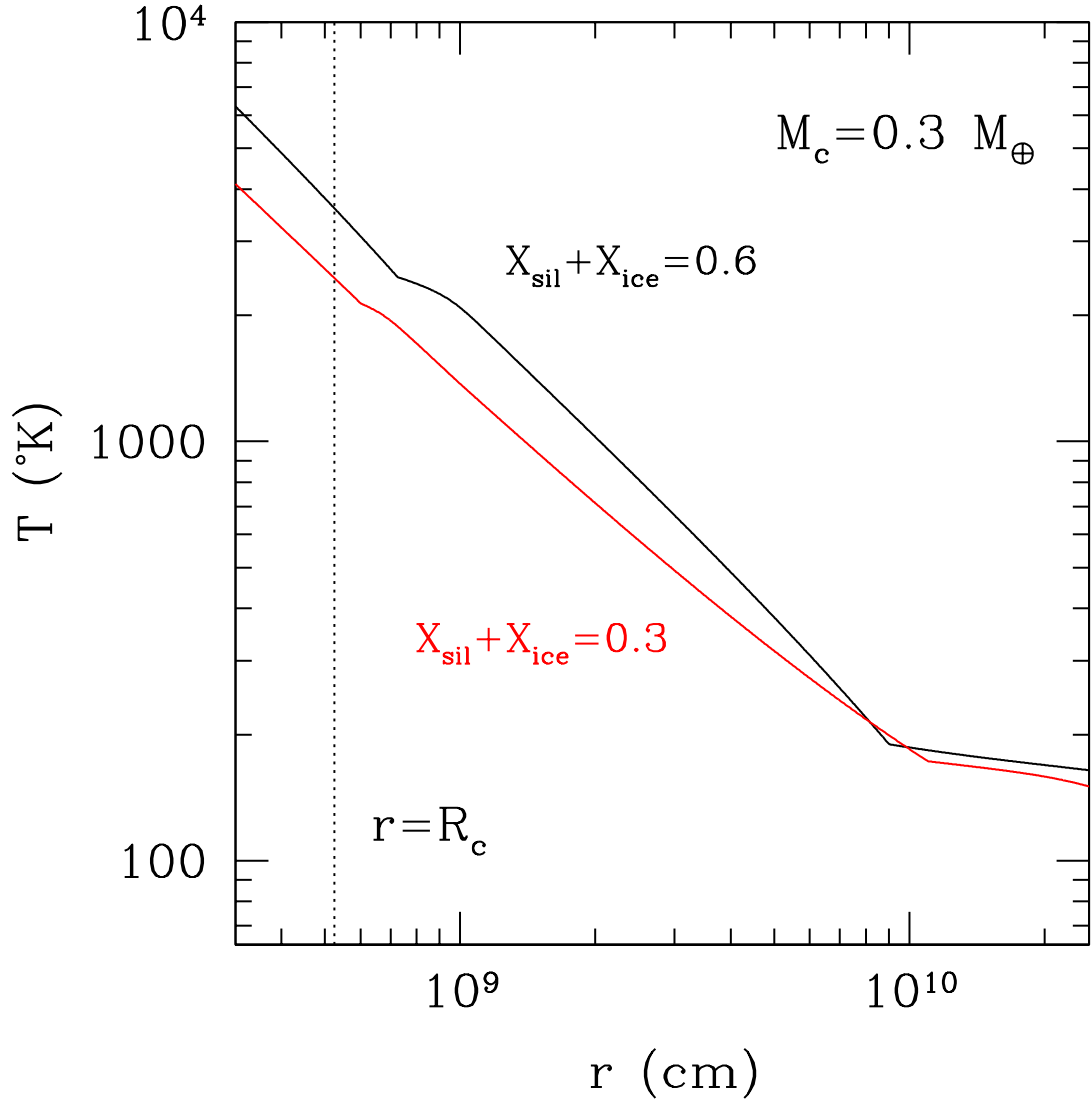}
        \vskip -1in
   \caption{Radial temperature profile of a non-gravitating, adiabatic atmosphere surrounding a point mass (0.3 $M_\oplus$) that is immersed in an asymptotically uniform medium of temperature 100 K and constant mass fraction of water ice and silicates, $X_{\rm ice} = X_{\rm sil}$.  Red and black curves correspond to $X_{\rm ice} + X_{\rm sil} = 0.3$ and 0.6.  
   Profile is calculated by combining Equation (\ref{eq:dtdp}) with the equation of hydrostatic equilibrium within the ice sublimation layer, and the analogous equation for silica within the silicate sublimation layer, as described in the text.  The temperature plateau associated with silicate sublimation is relatively mild and sits close to the inner computational boundary; hence we do not include the gas-silicate
heat exchange in our time-dependent envelope model.}
    \label{fig:ideal}
    \end{centering}
\end{figure}

Fully adiabatic profiles solving
Equations (\ref{eq:dpdr}), (\ref{eq:dtdr}), (\ref{eq:addvol}), (\ref{cp}), (\ref{eq:dtdp}),
and (\ref{eq:calP}) are shown in Figure \ref{fig:ideal} in the case of an ideal
gas ($\gamma = 1.4$ ratio of specific heats) EOS.
A broad plateau in the temperature in the outer envelope corresponds to the ice
sublimation layer.  The innermost envelope around temperature 1700 K also shows the effects of silicate
sublimation.  These are computed in an analogous manner by considering pure silica,
and considering the reaction ${\rm SiO}_2 \leftrightarrow {\rm SiO} + {1\over 2}{\rm O}_2$, implying
a mean molecular weight $\mu_{\rm sil} = 40m_u$ in the gas phase \citep{ts1988}.  
The sublimation energy is $\ell_{\rm sil} = 
1.6\times 10^{11}$ erg g$^{-1}$, and the vapor saturation pressure
$P_{\rm sil}^{\rm sat}(T) = 3.2\times 10^{14} e^{-(6\times 10^4{\rm K})/T}$ \citep{Krieger}.  

Our default envelope models do not include the effects of silicate sublimation on the settling of
dust across the core-envelope boundary.  In Section \ref{s:evol}, we consider how the imposition of
a sublimation barrier to dust settling changes the global mass flow of solids at low core luminosity.
One observes in Figure
\ref{fig:ideal} that a relatively weak temperature plateau is associated with silicate sublimation,
sitting just above the inner computational boundary.  
On that basis, we set aside the effects of heat exchange between solid and vapor silicates when
considering the effect of dust sublimation on mass transport.

\subsection{Radiative Transport}

The radiative temperature gradient is computed throughout the envelope as
\begin{equation}
\label{radgrad}
\nabla_{\text{rad}}=\frac{3 \kappa L_{\rm tot}}{64 \pi \sigma_{\rm SB} G M_c }\frac{P}{T^4}
\end{equation}
where 
\begin{equation}\label{eq:kappa}
\kappa = \kappa_g + \kappa_d
\end{equation}
is the net gas $+$ dust opacity.  The radiative luminosity has several contributions,
\begin{equation}
\label{eq:lumtot}
L_{\rm tot}(r)=
     L_{ p, \rm acc} + L_{\rm plan, acc} + L_{\rm 26Al} + L_{\rm KH} , 
\end{equation}
The accretion luminosity in 
pebbles (mass accretion rate $\dot M_p$) is approximated as follows.  A first contribution represents
the disintegration of the pebbles due to water ice sublimation;  the second represents the settling
of silicate pebble debris onto the core:
\begin{equation}
\label{lumacc}
L_{p,\rm acc}(r) = GM_c\dot{M}_p {\rm max}\left({1\over R_{\rm sub}} - {1\over r}, 0\right)
 + {GM_c\dot M_{\rm sett}(R_c)\over R_c},
\end{equation}
Here, $R_{\rm sub}$ is radius inside of which water ice disappears.  
The mean radial drift of the dust through the gas is dissipative, releasing the gravitational binding energy
of the pebble material,
\begin{equation}\label{eq:mdotsett}
\dot M_{\rm sett}(r) = 4\pi r^2 X_d \rho(r) \cdot \tau_{\rm stop} {GM_c\over r^2}.
\end{equation}
This quantity grows in magnitude toward the core, hence our approximation in Equation (\ref{lumacc}).
The accretion luminosity of larger planetesimals which can fully penetrate the envelope is
\begin{equation}
    L_{\rm plan, acc} = {GM_c \dot{M}_{\rm plan}\over R_c}.
\end{equation}

Finally $L_{\rm 26Al}$ is the luminosity emitted from the envelope due to the decay of $^{26}$Al
contained in dust within the envelope,
\begin{equation}
L_{\rm 26Al} = 1.5\times 10^{24} 
\left({M_d\over M_\oplus}\right) 
e^{-t/\tau_{\rm 26Al}} \quad {\rm erg~s^{-1}}.
\end{equation}
Here $\tau_{\rm 26Al} = 1.0$ Myr.
Both $L_{\rm 26Al}$ and 
the hydrostatic contraction luminosity $L_{\rm KH}$ of the core and inner envelope are
typically negligible compared with the two preceding luminosities, for the relevant
density ($\rho \gtrsim 1$ g cm$^{-3}$) and opacity ($\kappa \gtrsim 1$ cm$^2$ g$^{-1}$; \citealt{lee}).

The gas opacity $\kappa_g$ is represented following Equations (3)-(5) in \cite{freedman}. This 
prescription depends on the gas metallicity and is thus suitable for a water vapor-enriched atmosphere. One defines
\begin{equation}
\kappa_g = \kappa _ { \mathrm { lowP } } + \kappa _ { \mathrm { highP } }
\end{equation}
where
\begin{equation}
\begin{aligned} \log _ { 10 } (\kappa _ { \mathrm { lowP } }) = & c _ { 1 } \tan ^ { - 1 } \left( \log _ { 10 } T - c _ { 2 } \right) \\ & - \frac { c _ { 3 } } { \log _ { 10 } P + c _ { 4 } } e ^ { \left( \log _ { 10 } T - c _ { 5 } \right) ^ { 2 } } + c _ { 6 } m e t + c _ { 7 } \end{aligned}
\end{equation}
and
\begin{equation}
\begin{aligned} \log _ { 10 } (\kappa _ { \mathrm { highP } }) = & c _ { 8 } + c _ { 9 } \log _ { 10 } T \\ & + c _ { 10 } \left( \log _ { 10 } T \right) ^ { 2 } + \log _ { 10 } P \left( c _ { 11 } + c _ { 12 } \log _ { 10 } T \right) \\ & + c _ { 13 } m e t \left[ \frac { 1 } { 2 } + \frac { 1 } { \pi } \tan ^ { - 1 } \left( \frac { \log _ { 10 } T - 2.5 } { 0.2 } \right) \right], \end{aligned}
\end{equation}
where $c_{xx}$ are the coefficients shown in Table
\ref{tabop} and ``${met}$'' is the gas metallicity.

\begin{table}
\caption{Gas Opacity Coefficients}
\label{tabop}
\begin{center}

\begin{tabular}{lclccc}
\cline{1-6}
\multicolumn{3}{c}{For all T}           & \multicolumn{1}{l}{}    & \multicolumn{1}{l}{T \textless 800 K} & \multicolumn{1}{l}{T \textgreater 800 K} \\ \hline
c1         & \multicolumn{2}{c}{10.602}   & \multicolumn{1}{c}{c8}  & -14.051                                & 82.241                                    \\ 
c2         & \multicolumn{2}{c}{2.882}    & \multicolumn{1}{c}{c9}  & 3.055                                  & -55.456                                   \\ 
c3         & \multicolumn{2}{c}{6.09$\times 10^{-15}$} & \multicolumn{1}{c}{c10} & 0.024                                  & 8.754                                     \\ 
c4         & \multicolumn{2}{c}{2.954}    & \multicolumn{1}{c}{c11} & 1.877                                  & 0.7048                                    \\ 
c5         & \multicolumn{2}{c}{-2.526}   & \multicolumn{1}{c}{c12} & -0.445                                 & -0.0414                                   \\ 
c6         & \multicolumn{2}{c}{0.843}    & \multicolumn{1}{c}{c13} & 0.8321                                 & 0.8321                                    \\ 
c7         & \multicolumn{2}{c}{-5.490}   & ---                      & ---                                    & ---                                       \\ 
\cline{1-6}
\end{tabular}
\end{center}

\end{table}

The  grain opacity is taken to be the minimum of twice the geometric opacity and the Rosseland mean small-grain opacity
$\kappa_{d,\rm R}$ derived in Appendix \ref{s:kapgr},
\begin{equation}\label{eq:kapgr}
\kappa_d = \kappa_{\rm geom}Q.
\end{equation}
Here, $\kappa_{\rm geom} = 3X_d/4\rho_s a_d$ for spherical grains of radius $a_d$, material density $\rho_s$,
and mass $X_d$ per unit total mass, and
\begin{equation}\label{eq:Q}
  Q = {\rm min}(2, Q' \cdot 2\pi a_d / \lambda_{\rm max}), 
\end{equation}
where $\lambda_{\rm max}(T) \equiv hc/4.95\,k_{\rm B}T$ is the peak wavelength in the
Planck function.
The coefficient $Q' = 0.35$ for small silicate-carbon grains, but is a factor $\sim 1/(6-7)$ smaller when these
grains are coated with an icy mantle. 
Dust size $a_d$ is calculated self-consistently using Equations (\ref{eq:frag1}) and (\ref{eq:frag2}).

\subsection{Dust Transport}\label{s:transport}

The transport of small dust grains through the envelope is governed by the advection-diffusion equation {\citep{thoul}:}
\begin{equation}
\label{eq:trans}
\frac{\partial C}{\partial t} =
\frac{1}{(\rho_d + \rho_g)}\bnabla\cdot[D_{\rm tot} (\rho_g + \rho_d) \bnabla C] 
- \bnabla\cdot({\bf V}_{\rm sett} C) + {\dot\rho_d\over{\rho_d+\rho_g}}.
\end{equation}
Here, $C=\rho_d / (\rho_d + \rho_g)$ is the dust concentration, $D_{\rm tot}$ is the total effective diffusion coefficient, and the second term on the right-hand side represents
inward settling of the grains in the central gravitational field, ${\bf V}_{\rm sett} = -\tau_{\rm stop} g(r)\hat r = -\tau_{\rm stop} GM(r)\hat r/r^2$.
Although settling can dominate turbulent stirring as a source of differential grain velocities in the inner envelope, the bulk transport
of small grains is dominated by convection wherever it is present.  The source function $\dot\rho_p/(\rho_d+\rho_g)$
represents continuous pebble accretion, and therefore is centered at the pebble destruction radius as determined via
Equations (\ref{eq:pebdes1}) and (\ref{eq:pebdes2}).  We convolve the source term with a narrow normal distribution to enhance numerical stability.     

We solve Equation (\ref{eq:trans}) numerically using \texttt{FiPy}: A Finite Volume PDE Solver Using Python\footnote{https://www.ctcms.nist.gov/fipy/} \citep{guyer}. Although $D_{\rm tot}$ in principle receives a contribution from molecular diffusion, eddy diffusion
completely dominates in convective zones, and the advection term dominates in radiative zones. We set $D_{\rm tot} = D_{\rm con} + D_{\rm mol}$, where
\begin{equation}\label{eq:dtot}
\begin{split}
D_{\rm con} & = (1+{\rm St_d}^2)^{-1} H V_{\rm con} \\
& = (1+{\rm St_d}^2)^{-1} H \sqrt{gH(\nabla - \nabla_{\rm ad})},  
\end{split}
\end{equation}
and
\begin{equation}\label{eq:dmol}
\begin{split}
D_{\rm mol} \sim c_s \ell_g .
\end{split}
\end{equation}
Here, $H$ is the pressure scale height (corresponding to a mixing length equal to $H$), and ${\rm St_d}$ is the dimensionless dust Stokes number ${\rm St_d} = V_{\rm con}\tau_{\rm stop}/H$, with $\tau_{\rm stop}$ defined in Equation (\ref{eq:ts}).

The superadiabatic gradient $\nabla - \nabla_{\rm ad}$ is obtained from a mixing length model of convection, 
following the prescription of \cite{cox}.  One first defines the ratio of convective to radiative conductivity,
\begin{equation}
\label{cond}
A = \frac{Q^{1/2}C_p \kappa g \rho^{5/2}H ^2}{12\sqrt{2}acP^{1/2}T^3},
\end{equation}   
{where $Q = 1 - (\partial \ln \mu/ \partial \ln T)_P$} {with $\mu$ the mean molecular weight,} and $C_p$ the heat capacity at constant pressure, as derived
from the EoS. We then define the quantity
\begin{equation}
B = \bigg[\frac{A^2}{a_0}(\nabla_{\rm rad} - \nabla_{\rm ad})\bigg]^{1/3}; \quad\quad a_0 \equiv {9\over 4}.
\end{equation}
Finally, the convective efficiency
\begin{equation}
\zeta = \frac{\nabla_{\rm rad} - \nabla}{\nabla_{\rm rad} - \nabla_{\rm ad}}
\end{equation}
is obtained by solving the polynomial equation
\begin{equation}
\zeta^{1/3} + B\zeta^{2/3} + a_0B^2\zeta - a_0B^2 = 0.
\end{equation}
Once $\zeta$ is known, one immediately obtains $\nabla - \nabla_{\rm ad}$ 
and the convective velocity from Equation (\ref{eq:dtot}). 

Note that in the mixing length model, $V_{\rm con} \propto \sqrt{\nabla - \nabla_e}$, where $\nabla_e$ is the 
temperature gradient of the rising element. This factor can be approximated as $\sqrt{\nabla - \nabla_{\rm ad}}$
when the element experiences minimal heat loss by radiation, corresponding to efficient adiabatic convection. 
The applicability of this condition, and of the corresponding simple approximation $V_{\rm con}^3 \simeq 
L_{\rm con}/\rho$ (implying comparable contributions of the kinetic energy and enthalpy perturbation to the 
energy flux) is reviewed in Appendix \ref{s:coneff}.

The dust mass fraction in the outer envelope is $X_d = X_{\rm ice} + X_{\rm sil}$.
The  ice mass fraction is easily derived by subtracting the saturation water vapor density
from the net solid$+$vapor water mass density, 
$X_{\rm ice} = X_{\rm H_2O} - P_{\rm H_20}^{\rm sat}(T) \mu_{\rm H_2O}/\kB T\rho$. 
The analogous procedure is used for silicates when we construct alternative models
with dust sublimation in the inner envelope.

\begin{table}
\caption{Adjustable Parameters of the Model} 
\begin{center}
{\begin{tabular}{lcccc}
\hline
\noalign{\smallskip}
Parameter			& value			& note	\\
\hline
$\dot{M}_{p}$ & $10^{-7}$, $10^{-6}$ $M_\oplus$ yr$^{-1}$ &	pebble accretion rate \\
$\dot{M}_{\rm plan}$ & 0, $10^{-7}$  $M_\oplus$ yr$^{-1}$&	 planetesimal accretion rate \\
$M_c$ & $0.3\,M_\oplus$ & initial protoplanet (core) mass \\
$R_c$ & 5200 km  & core radius (inner boundary)\\
$\rho_{g,\rm disk}$ & 10$^{-11}$ g cm$^{-3}$ &	 disk gas density \\
$T_{\rm disk}$ & 100 K &	disk temperature \\
$R_{\rm B}$ & $3.3\times 10^{10}$ cm  &  Bondi radius ($\sim$ outer boundary) \\
$a_{p}$ & 0.1 cm &	 radius of injected pebbles \\
$a_{d,0}$ & 1 $\mu$m &	 dust radius at $t=0$\\
$X_{d,\rm disk}$ & $0.01$ &	 disk metallicity\\
$X_{\rm ice} = X_{\rm sil}$ & &  mass fractions in injected dust/pebbles \\
$V_f$ & $10^2/10^3$ cm s$^{-1}$ &	 dust/ice fragmentation speed \\

\hline	
\end{tabular}}
\label{t4}
\end{center}
\tablecomments{The main suite of models assumes $M_c = 0.3\,M_\oplus$, 
$\dot M_p = 10^{-7}$, $10^{-6}\,M_\oplus$ yr$^{-1}$,
standard pebble composition $X_{\rm ice} = X_{\rm sil}$, 
and variable planetesimal accretion rate $\dot{M}_{\rm plan}$. 
For comparison, two models with ice-free pebbles are also produced.}
\end{table}

\subsection{Initial Conditions and Boundary Conditions}\label{s:boundary}

The envelope is initialized as a solution to the hydrostatic and radiative Equations 
(\ref{eq:dpdr})-(\ref{eq:addvol}) and (\ref{eq:dtdp})-(\ref{eq:kappa}) with $M_c = 0.3\,M_\oplus$,
 uniform dust abundance $X_d = X_{d,\rm disk} = 0.01$, $X_{\rm ice} = X_{\rm sil}= X_d/2$, 
and grain size $a_d = 1$ $\mu$m (Table \ref{t4}).

Constant conditions $\rho_g = \rho_{g,\rm disk} = 10^{-11}$ g cm$^{-3}$, $T = T_{\rm disk} = 100$ K and 
$X_d = X_{d,\rm disk}$ are maintained at the outer computational boundary $R_{\rm max} 
= 4\times 10^{10}(M_c/0.3~M_\oplus)$ cm $\gtrsim R_{\rm B}$.  Grains produced by the destruction of larger pebbles
are allowed to flow outward across this boundary and, 
since the external grain density is held fixed, this outflow
represents a permanent loss from the envelope to the disk.

The inner computational boundary sits at the core radius $R_c = 5.2\times 10^8$ cm.  Here an
open boundary condition is also applied for silicate grains, meaning that settling grains 
that reach the core are accreted and removed permanently from the envelope. In our default configuration, 
we do not allow silicates to vaporize in the inner envelope even when this is 
thermodynamically possible.  The effect of an inner silicate sublimation barrier (Section \ref{s:rain})
on the envelope mass flow is considered as an alternative case.

As regards the outer boundary condition on $X_d$, two arguments suggest that
the equilibrium metal concentration in the envelope
is relatively insensitive to the ambient metallicity.  The high dust abundance that is reached in our 
envelope models (10-30 times Solar) is mainly due to the weak coupling of pebbles to the gas around 
the Bondi radius, which allows the solids to concentrate with respect to the H$_2$/He gas.  An outflow of
metal-enriched gas will increase the metallicity of the co-orbital region in the surrounding disk, and
so the feedback on the equilibrium envelope metallicity should be considered.

First, we tested the consequences of a free-flow outer boundary condition, by extending the computational domain
to well beyond the Bondi radius.  Dust transport was tracked in this outer zone, and convection allowed to develop
there.  We found the results inside radius $R_{\rm B}$ to differ minimally from those obtained 
from the default boundary conditions. 
Second, one can estimate analytically the sensitivity of the equilibrium $X_{d,\rm env}$ in the envelope
to the ambient $X_{d,\rm disk}$.  Given a fixed pebble accretion rate, and a steady state relation between 
solid accretion and excretion rates,
\begin{equation}
\dot M_p \simeq 4\pi R_{\rm B}^2 V_{\rm con}(R_{\rm B}) \rho_g(R_{\rm B}) (X_{d,\rm env} - X_{d,\rm disk}),
\end{equation}
one sees that it is difference $X_{d,\rm env}-X_{d,\rm disk}$ in dust abundance between envelope and disk that is
constrained.   If we were 
arbitrarily to raise $X_{d,\rm disk}$ from the assumed Solar abundance up to the value that is reached in a given envelope
model, then the new envelope metallicity would be a factor of 2 larger.
In practice, the dust component of the ejected gas must be diluted 
by mixing with lower-metallicity disk gas, or by the reassembly of small grains into pebbles.

The entropy of the inner envelope does not initially match the ambient disk entropy,
but continuing solid accretion combined with secular dust accumulation in the
envelope removes this imbalance in entropy.  This result can be seen both semi-analytically
and numerically. 
The steady envelope solution presented in Section \ref{s:2layer} has a uniform
entropy in the inner convection zone that is lower than the entropy of the ambient medium.
(In this model, the central radiation source that drives the convection represents deep planetesimal impact heating.)
These inner and outer zones are connected by a layer with radially increasing entropy (Figure
\ref{fig:radprof}).  As the
envelope is loading with an increasing amount of dust (here produced by pebble destruction),
the entropy of the convection zone rises.  Finally the entropy is nearly uniform in
the fully convective state with the parameter $\varepsilon_{\rm rad} \gtrsim 0.3$.  Only then is
the pebble debris recycled back into the disk.  

\begin{figure}
\begin{centering}
        \includegraphics[scale=0.40]{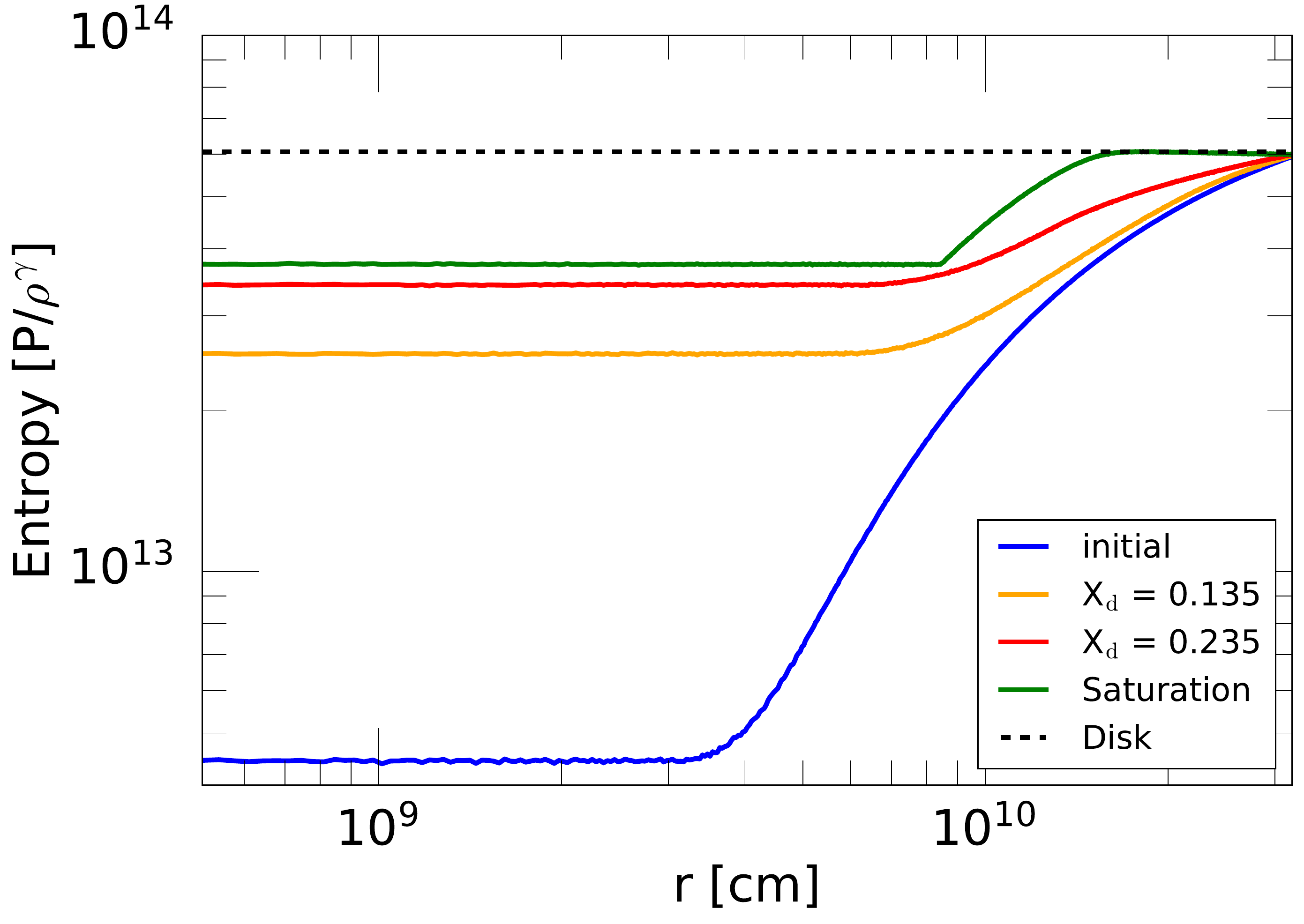}
   \caption{ Generalized entropy profile of the H$_2$/He gas in the envelope as a function of increasing dust abundance (thus time).
   Case shown here is the high-luminosity model with $\dot{M}_p = 10^{-6}\,M_\oplus$ yr$^{-1}$
   and $\dot M_{\rm plan} = 0$.   The entropy increases with time (and envelope
   metallicity) until the envelope is fully convective and its entropy matches the disk entropy.
   In the two lowest-metallicity snapshots (red and orange) the bend at $\sim 6-7\times 10^9$ cm represents the boundary between
   the inner convection zone and the outer radiative layer.
In the two highest-metallicity snapshots (green and blue) most of the rise outside $10^{10}$ cm is due to heat exchange with
sublimating ice;  at this stage, the RCB has moved out to the outer computational boundary.}
    \label{fig:entropy}
    \end{centering}
\end{figure}

Figure \ref{fig:entropy} shows the time evolution of the gas `entropy' function $P/\rho_g^{7/5}$ for our
nominal time-dependent model with icy pebbles (see Section \ref{s:results}) and heating driven
  by the accretion of pebble debris.
The gas entropy is flat in the inner convection zone, and rises through the outer envelope
due to a combination of radiative transport and also the extraction of heat by ice sublimation. 
  As time progresses, the radiative 
layer shrinks and the convective plateau grows.  The outer entropy bump does not entirely disappear
in the fully convective state, due to the continuing sublimation of ice.
See Section \ref{s:evol} for further details.

In sum, we do not expect that a ``buoyancy barrier'' should be present when envelope heating overwhelms cooling. The numerical model constructed
by \cite{kurokawa} does not include planetesimal heating or dust production. 
A fuller
exploration of the interplay between heating, cooling, and hydrodynamic expulsion will require incorporating these effects into global simulations.

\section{Results}\label{s:results}

We now describe the time-dependent behavior of the accreted atmosphere, while it remains in thermal contact with
the disk.  We ran simulations using the model described in Sections \ref{s:struc}-\ref{s:transport}
and the boundary conditions summarized in Section \ref{s:boundary}, with a uniform set of parameters
summarized in Table \ref{t4}.  Our focus is on the early growth of a core, $M_c = 0.3\,M_\oplus$.
The default model assumes pebbles of mixed silicate/ice composition ($X_{\rm ice}=X_{\rm sil}$), which we compare with
the case of dry pebbles ($X_{\rm ice} = 0$).  The chosen core mass is just above the minimum mass that accretes a
high-entropy H$_2$/He envelope hot enough to completely sublimate silicates in a finite layer outside the core
(when the core itself is silicate-rich).  Our default model assumes that the core-envelope boundary is {\it permeable}
to silicate dust.  We also compare the results for envelope and core mass accumulation with an alternative
situation where silicate sublimation prevents growth of the core by silicate `rain' (see Section \ref{s:rain}
for motivation).

The ambient disk temperature and H$_2$/He mass density are taken to be $T_{\rm disk} = 100$ K and
$\rho_{g,\rm disk} = 10^{-11}$ g~cm$^{-3}$ (Table \ref{t4}), corresponding to moderate coupling of the
pebbles to the gas near the Bondi radius (Equation (\ref{eq:rhodisk})).  The outer and inner computational
boundaries sit at $R_{\rm max} = 4\times 10^{10}$ cm $\gtrsim R_{\rm B}$ and $R_c = 5.2\times10^{8}$ cm.
Gas density and temperature are matched to the disk values at radius $R_{\rm max}$, and pebbles
are introduced into the envelope in the manner described in Section \ref{s:model}.  

A principal goal is to determine the minimum luminosity (core accretion rate) that is needed to drive
refluxing of the small particles across the Bondi radius.   We fix the pebble accretion rate $\dot M_p$ at a
conservative values of $10^{-7}$, $10^{-6}$ M$_\oplus$ yr$^{-1}$ (the larger quantity is 
needed to e-fold the core mass up to $\sim 10\,M_\oplus$ in a Myr) 
and test two values (0 and $10^{-7}$ M$_\oplus$ yr$^{-1}$) for the planetesimal accretion rate 
$\dot{M}_{\rm plan}$.  The core is
assumed to have a rocky composition; to obtain the same heating rate by accretion 
onto a rock-ice core with mean density $\widetilde\rho_c \times 3$~g~cm$^{-3}$, the 
accretion rate would need to rise by a modest factor $ 1.6 (\widetilde\rho_c/0.5)^{-2/3}$.

\begin{figure}
 \epsscale{0.8}
 \plotone{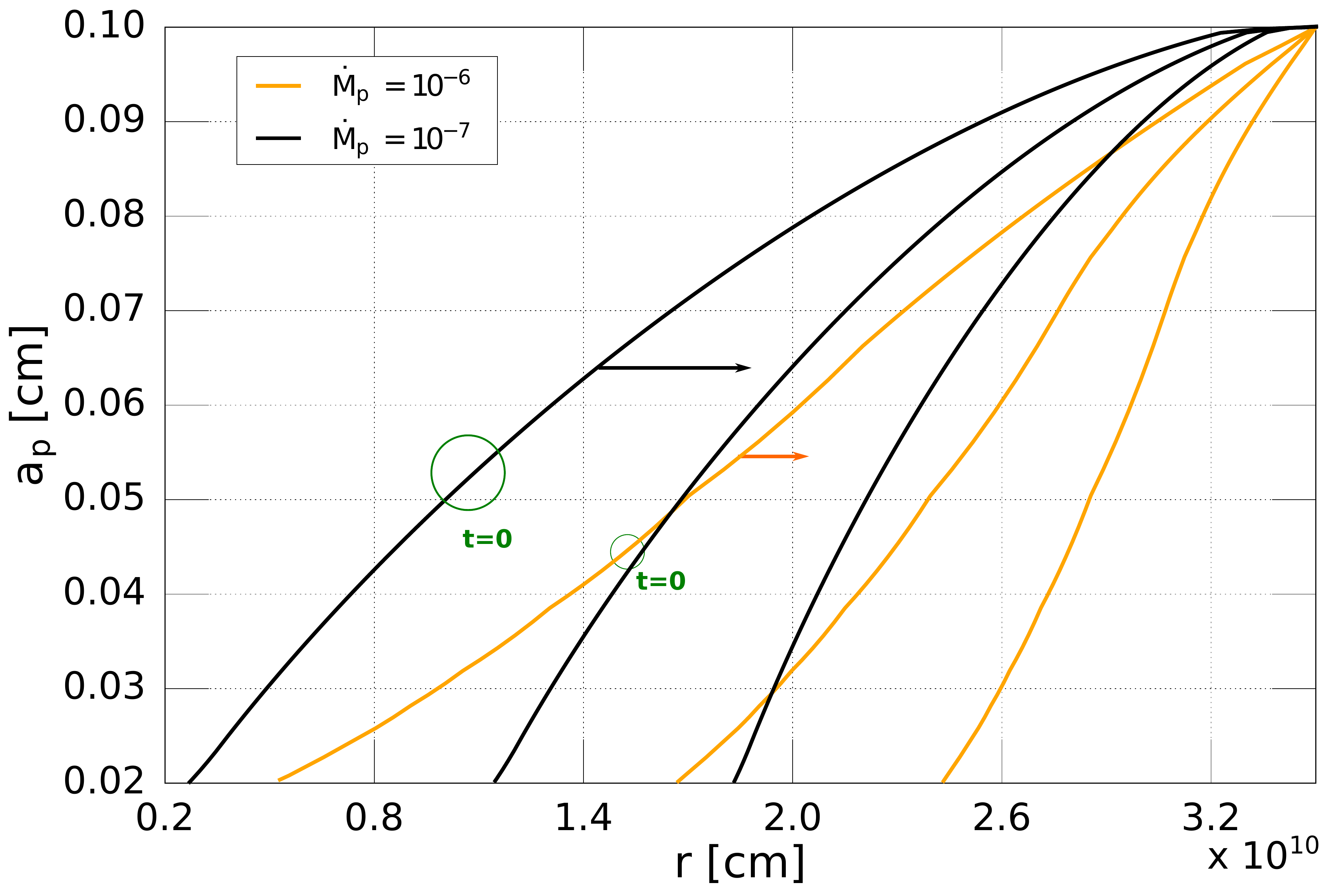}
  \caption{Trajectories of accreted icy pebbles with initial radius 0.1 cm as they are eroded by sandblasting (Equation 
(\ref{eq:pebdes2})) in the outer envelope. Curves marked with green circles represent pebble trajectories near
the start of the simulation. At later times, both sets of trajectories have shifted to the right, reaching a quasi-steady state,
as the density of small background particles builds up.
When $\dot M_p = 10^{-7}\,M_\oplus$ yr$^{-1}$ ($\dot M_{\rm plan} = 0$), the pebbles are destroyed in the outer radiative layer;
when $\dot M_p = 10^{-6}\,M_\oplus$ yr$^{-1}$, they are destroyed in the convective layer,
which in this case  has expanded to the outer computational boundary.}
     \label{fig:pebble}
\end{figure}

\subsection{Sandblasting and Sublimation of Pebbles}

Infalling pebbles are rapidly eroded by collisions with small ambient grains and by sublimation, as is seen in Figure
\ref{fig:pebble}.  After $X_d$ builds up in the outer envelope, destruction is almost complete a small distance
inside the outer boundary of the computational domain.  
We show two cases, one in which the outer envelope remains radiative ($\dot{M}_p$ = $10^{-7} M_\oplus$ yr$^{-1}$), 
and one in which convection extends beyond the Bondi radius ($\dot{M}_p = 10^{-6} M_\oplus$ yr$^{-1}$).

One see that, although destruction is faster in the presence of convection, secular drift in the central gravitational field is
still strong enough to eliminate the pebbles quickly in the outer radiative part of a low-luminosity envelope.
Pebble debris settles rapidly toward the inner convective layer:  at the fragmentation speed $V_f \sim 10$ m s$^{-1}$ for icy
particles, this takes a month or so, and for silicate particles about a year.  The dust opacity in the outer envelope
saturates over this timescale (Section \ref{s:evol}).

\subsection{Thermal and Hydrostatic Structure of the Envelope}

We now analyze the radial structure of the accreted envelope.  Two effects produce a temperature plateau in the outer
envelope:  (i) hydrostatic equilibrium in a radiative atmosphere with outer temperature buffered close to the disk temperature 
by dust-gas heat exchange
(see Figure \ref{fig:radprof}); and (ii) sublimation of water ice, which limits the growth of the gas temperature above the 
water sublimation temperature.  

\begin{figure}
\epsscale{1.1}
\plotone{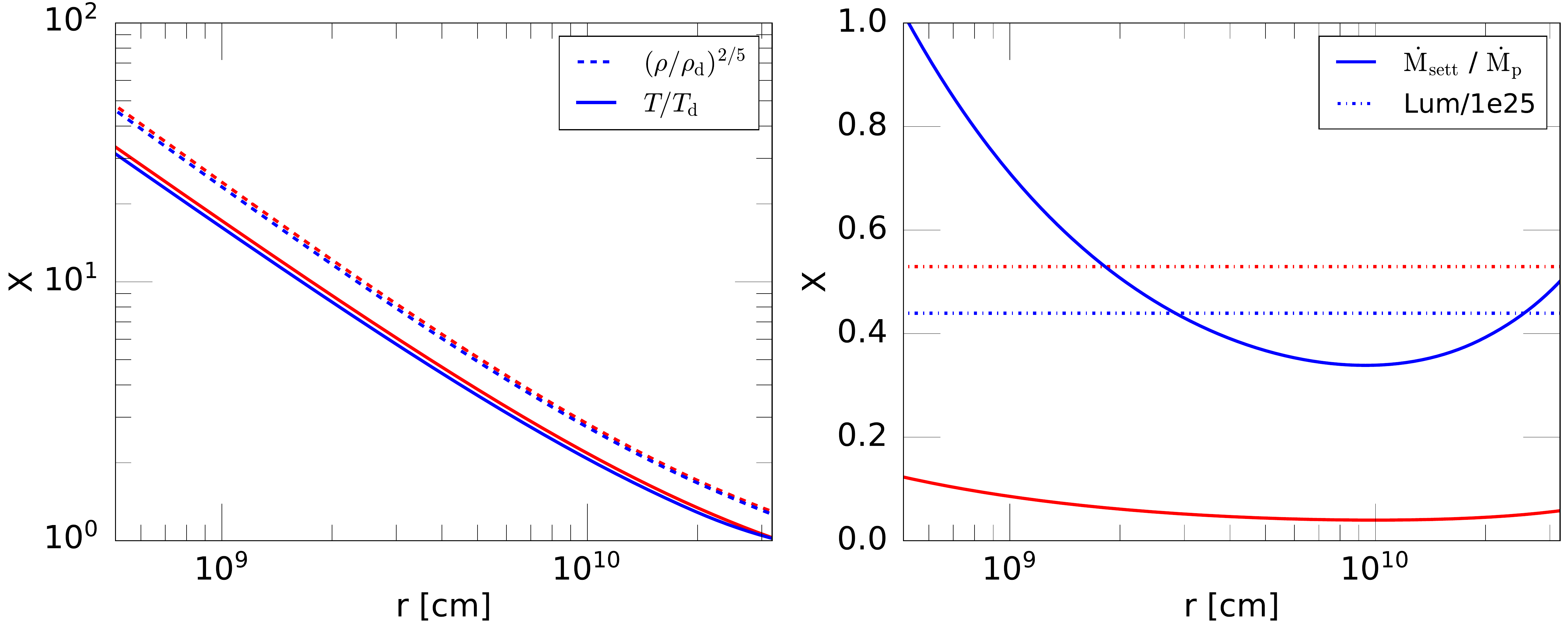}
   \caption{Profiles of the envelope accreted around a $0.3\,M_\oplus$ core,
   once the models have reached a steady state at $t \sim 10^3$ yr.   Here the accreted pebbles are ice free. 
   Blue curves show the low-$\dot M_p$ model with $\dot M_p = 10^{-7}\,M_\oplus$ yr$^{-1}$,
   $\dot M_{\rm plan} = 0$ and steady-state $X_d = 0.41$.  Red
   curves the high-luminosity model with $\dot M_p = 10^{-6}\,M_\oplus$ yr$^{-1}$
   and $X_d = 0.44$.  Left panel:  temperature (solid lines) and 
   total gas$+$dust density (dashed lines).  The envelope is fully convective in the high-luminosity
   model, but a radiative layer remains at low luminosity.  
   Right panel: the radius-dependent dust settling rate (Equation (\ref{eq:mdotsett}), solid curves)
   normalized to pebble accretion rate;  convective$+$radiative luminosity (dash-dot curves) in units
   of $10^{25}$ erg s$^{-1}$.
    All the pebble debris is accreted onto the core in the low-$\dot M_p$ model, but around
    $10\%$ in the fully convective state, with the remainder refluxing back into the disk.}
       \label{fig:tempc}
\end{figure}

The simplest case of nearly ice-free pebbles ($X_{\rm ice} = 10^{-4}$) is shown in Figure \ref{fig:tempc}, for
both low and high planetesimal accretion rates.   
The right panel shows the settling rate of dust embedded in the atmosphere due to the d.c. component of the
radial drift (Equation \ref{eq:mdotsett}).
In the low-$\dot M_p$ model (which does not produce refluxing of dust across the Bondi radius), this 
is an appreciable fraction of the imposed pebble accretion rate.  At
the core boundary there is a balance between pebble accretion and transmission of dust.
By contrast, when the accretion rate is high enough to drive dust refluxing (and limit the growth of
the envelope), the accretion rate of silicate dust onto the core is about $0.1\,\dot M_p$.

\vskip .1in

\begin{figure}
\epsscale{1.1}
\plotone{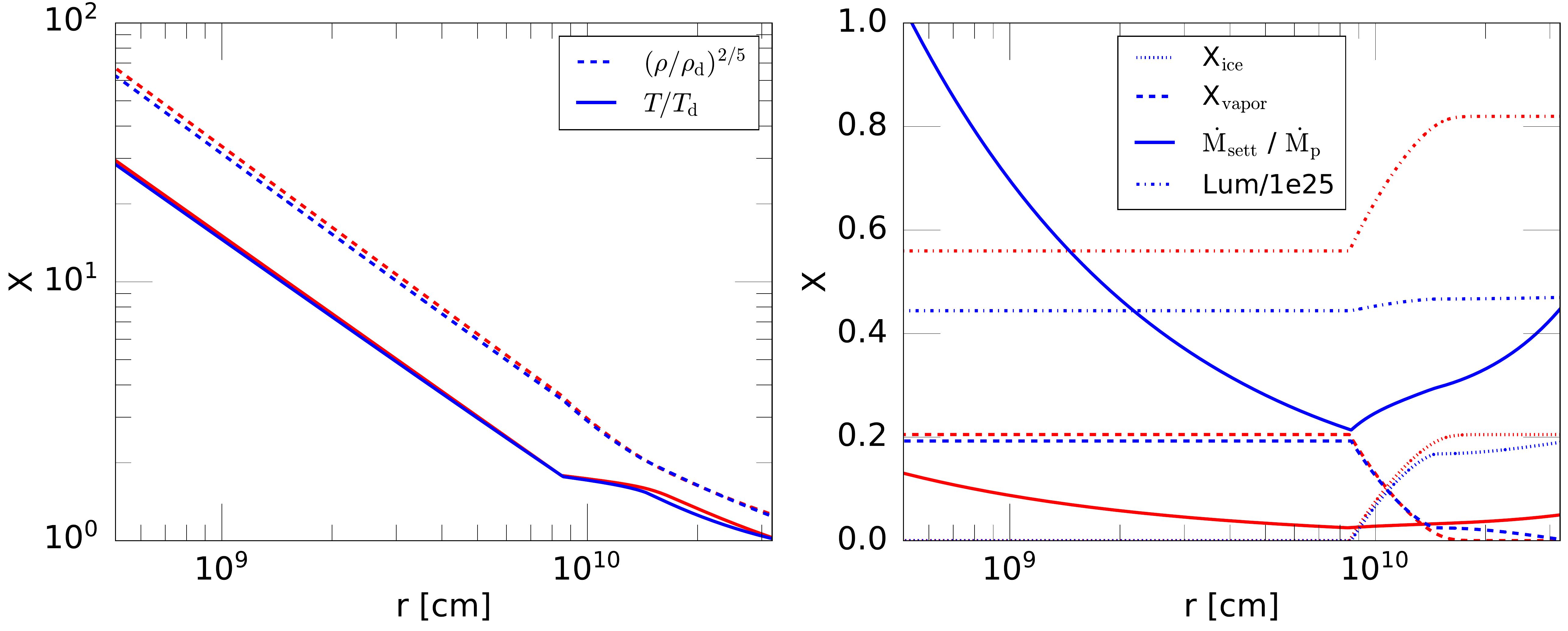}
   \caption{Repeat of Figure \ref{fig:tempc}, but now for icy pebbles with $X_{\rm ice} = X_{\rm sil}$. 
   Blue/red curves represent the low/high $\dot M_p$ models.  Now
    the outer temperature plateau is more extended, due to the heat exchange between the pebbles and 
the vapor phase. The right panel also shows the mass fraction $X_{\rm ice}$ of water ice in pebble debris 
(dotted curves) and the mass fraction $X_{\rm vapor}$ of water vapor from sublimated pebbles (dashed
curves).}
    \label{fig:temp}
\end{figure}

The core
accretion rate $\dot M_{\rm sett}(R_c)$ saturates at $\sim 1\times 10^{-7}\,M_\oplus$ yr$^{-1}$ for intermediate
values of the pebble accretion rate (see Table \ref{tbind}).   Convection reaches the Bondi radius when $\dot M_p$ slightly exceeds
this value, with most of the pebble debris continuing to accrete onto the core until $\dot M_p$ rises above
$\sim 2\times 10^{-7}\,M_\oplus$ yr$^{-1}$.  We therefore derive a core mass e-folding time due to pebble accretion
no shorter than $\sim 3$ Myr:  trapping of pebbles at higher rates implies higher rates
of refluxing but no faster core growth.

Turning next to the case of icy pebbles ($X_{\rm ice} = X_{\rm sil}$), one sees in Figure \ref{fig:temp} that
the temperature profile shows an outer plateau at both low and high accretion luminosities. 
The inner density is about a factor 2 higher and the dust settling rate is similar to the case of ice-free pebbles.

\begin{figure}
\epsscale{0.75}
\plotone{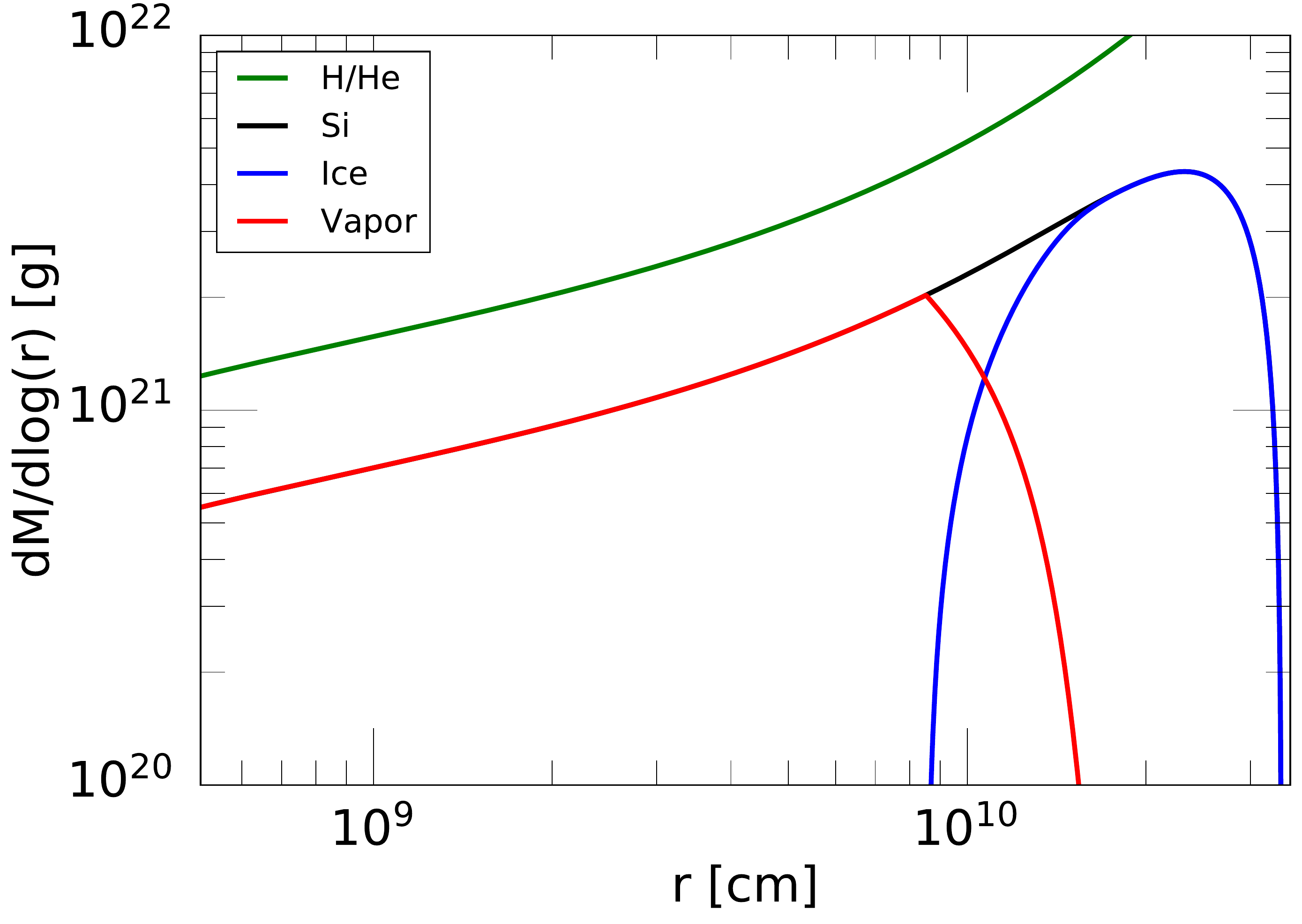}
   \caption{Mass profiles of H$_2$/He (green curve), silicate dust (black curve), water ice (blue curve)
   and water vapor (red curve) in the saturated high-luminosity model ($\dot M_p = 10^{-6}\,M_\oplus$ yr$^{-1}$,
   $\dot M_{\rm plan} = 0$) at $t \sim 10^3$ yr.
   The injected pebbles have composition $X_{\rm sil} = X_{\rm ice}$, hence the silicate dust and 
   water vapor curves overlap inside $\sim 10^{10}$ cm.}  
    \label{fig:solidmasses}
\end{figure}

The spatial distributions of H$_2$/He, silicate dust, ice, and water vapor mass are shown in Figure 
\ref{fig:solidmasses} for the higher $\dot M_p$ model.  Ice is present beyond a radius 
$R_{\rm sub} = 9\times 10^9$ cm,
which marks the termination of the temperature plateau seen in Figure \ref{fig:temp}.  

\begin{figure}
\epsscale{0.75}
\plotone{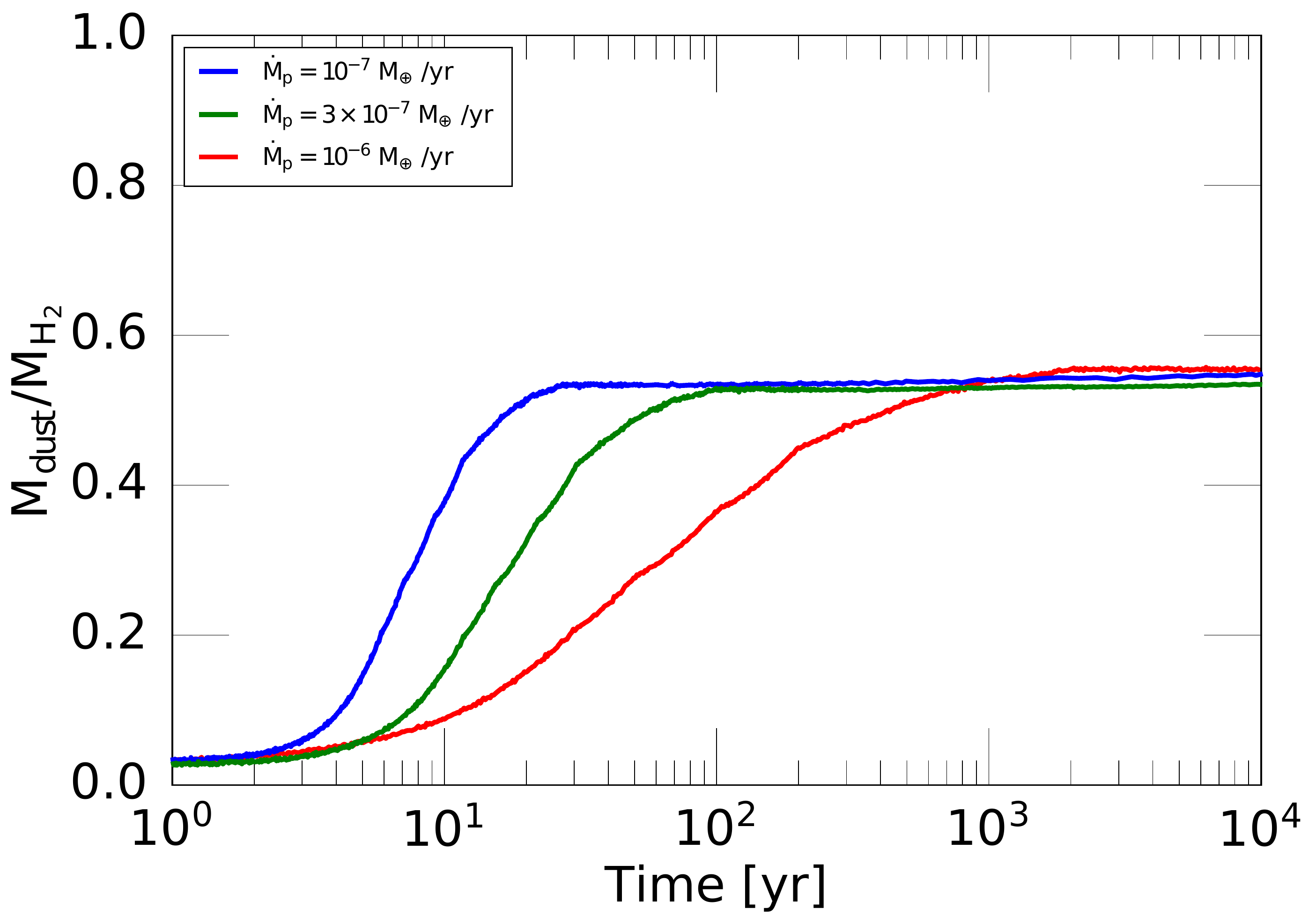}
   \caption{Ratio of silicate$+$ice dust mass to H$_2$/He mass in the envelope as a function of time, for the two fiducial
   models (blue curve: low $\dot M_p$; red curve: high $\dot M_p$) and also an intermediate model
   with $\dot M_p = 3\times 10^{-7}\,M_\oplus$ yr$^{-1}$ (green curve).  The envelope becomes fully convective
   in the two highest luminosity models, refluxing $70-90\%$ of the accreted pebble material back
   into the disk.}
    \label{fig:massratio}
\end{figure}

\begin{figure}
\epsscale{0.75}
\plotone{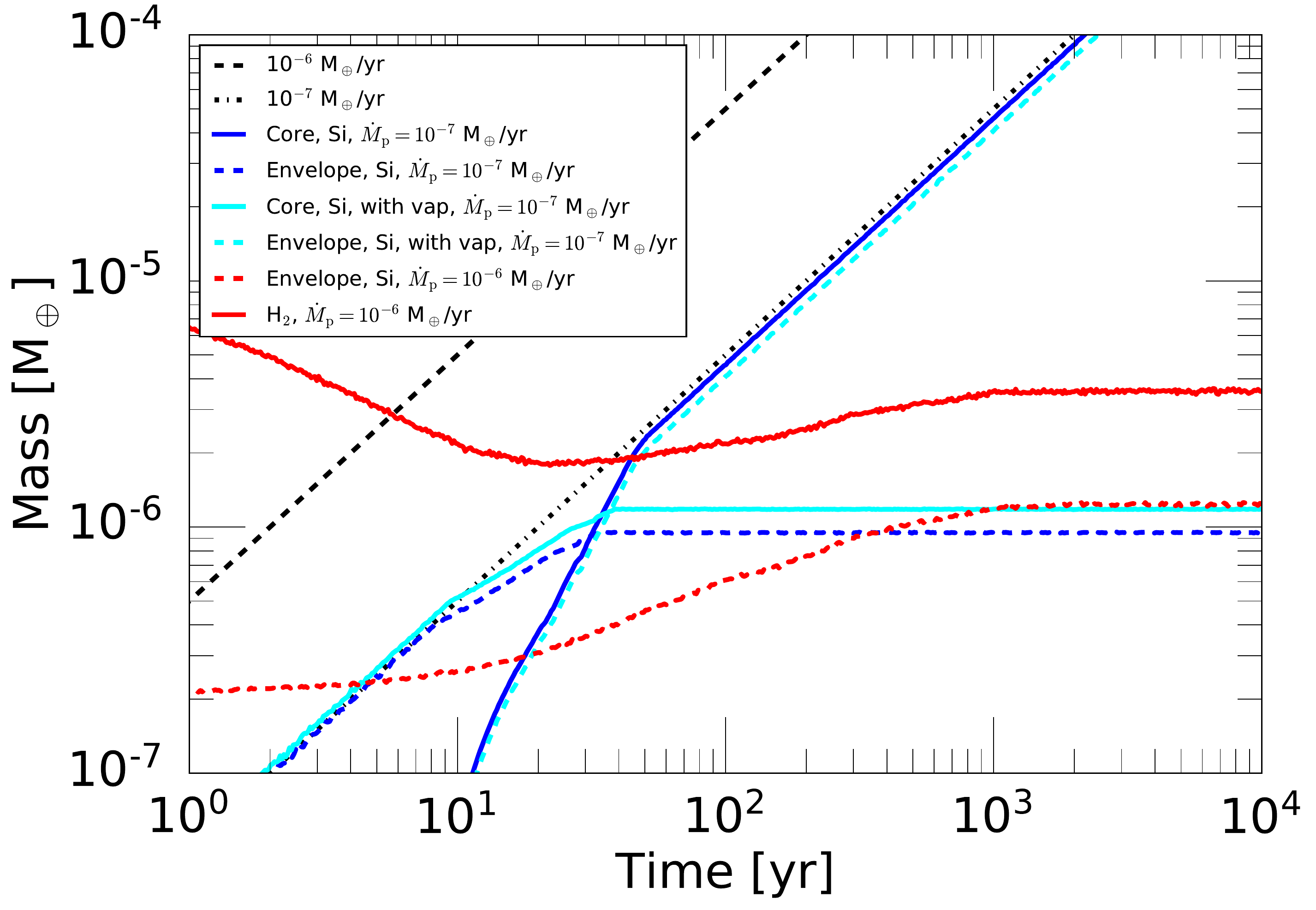}
   \caption{Time evolution of silicate core mass (solid curves) and envelope dust mass (short-dashed curves). 
   Blue curves show the default
   low-$\dot M_p$ model, in which silicate dust can accrete onto the core through the inner computational
   boundary.  Cyan curves show an alternative low-$\dot M_p$ model in which solid accretion onto the core is
   suppressed by the sublimation of silicate dust in the inner 1-2 scale heights.  When silicate dust
   can accrete, the core grows at the imposed pebble accretion rate (compare solid blue and black curves).
   On the other hand, when sublimation suppresses dust accretion (but the luminosity is low), the pebble debris
   is stored in the envelope (cyan dashed curve) and the core mass quickly saturates (cyan solid curve).   
   Red curves show the high-luminosity model ($\dot M_p = 10^{-6}\,M_\oplus$ yr$^{-1}$), in which
   dust slowly builds up
   in the envelope as the RCB oscillates around the Bondi radius, until the model reaches saturation after 
   $\sim 10^3$ yr. Beyond this point, all but $\sim 10\%$ of the accreted solid mass is recycled back into the
   disk; the core mass grows slowly because silicate sublimation is turned off, and the equilibrium 
   envelope H$_2$/He mass (red long-dashed curve) is about twice the envelope dust mass. }
       \label{fig:dustnew}
\end{figure}

\begin{figure}
\epsscale{0.75}
\plotone{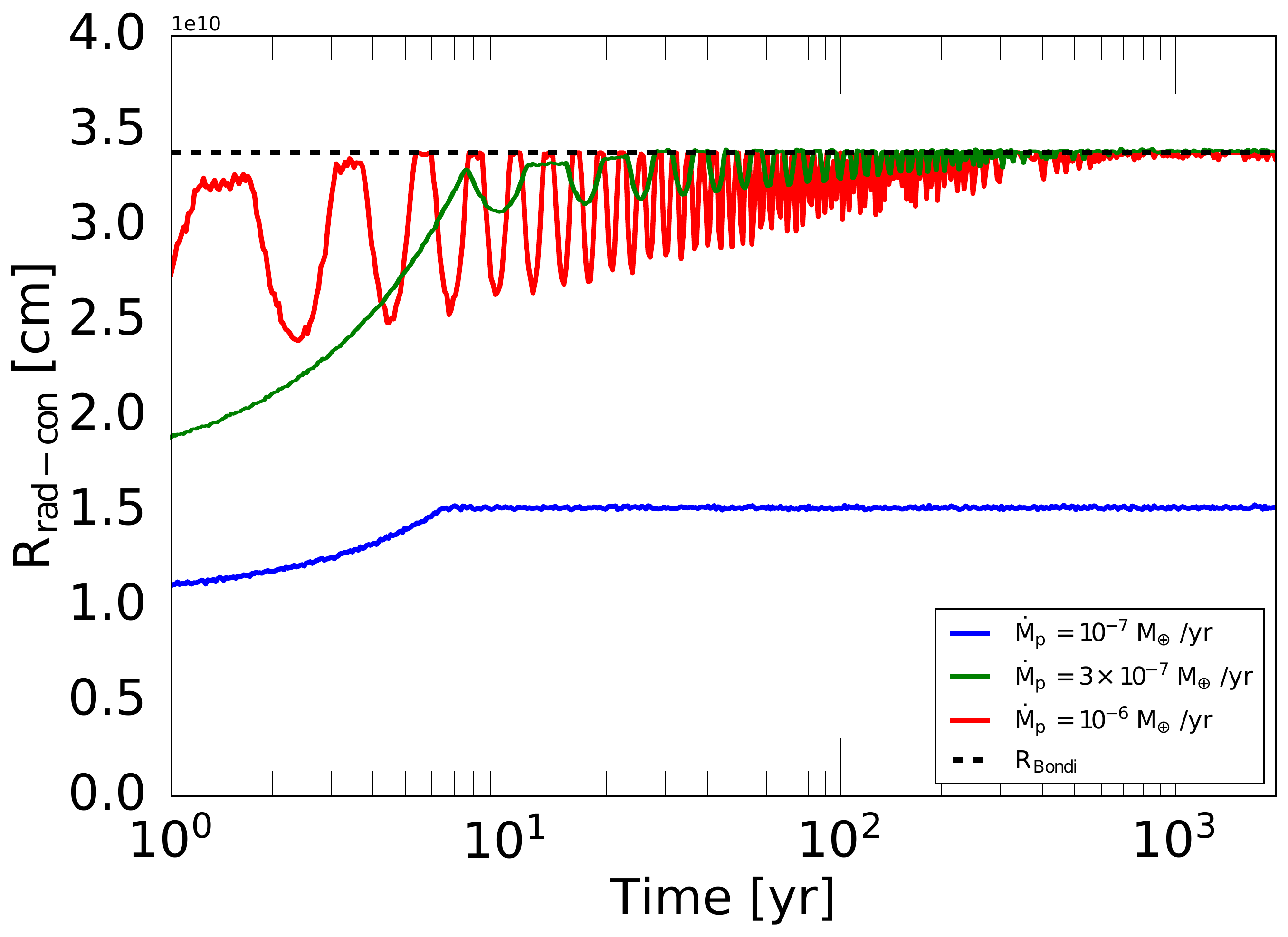}
   \caption{Radial position of the radiative-convective boundary in the envelope as a function of time, for
    the models shown in Figure \ref{fig:massratio}.}
    \label{fig:main}
\end{figure}

\begin{figure}
\epsscale{0.8}
\plotone{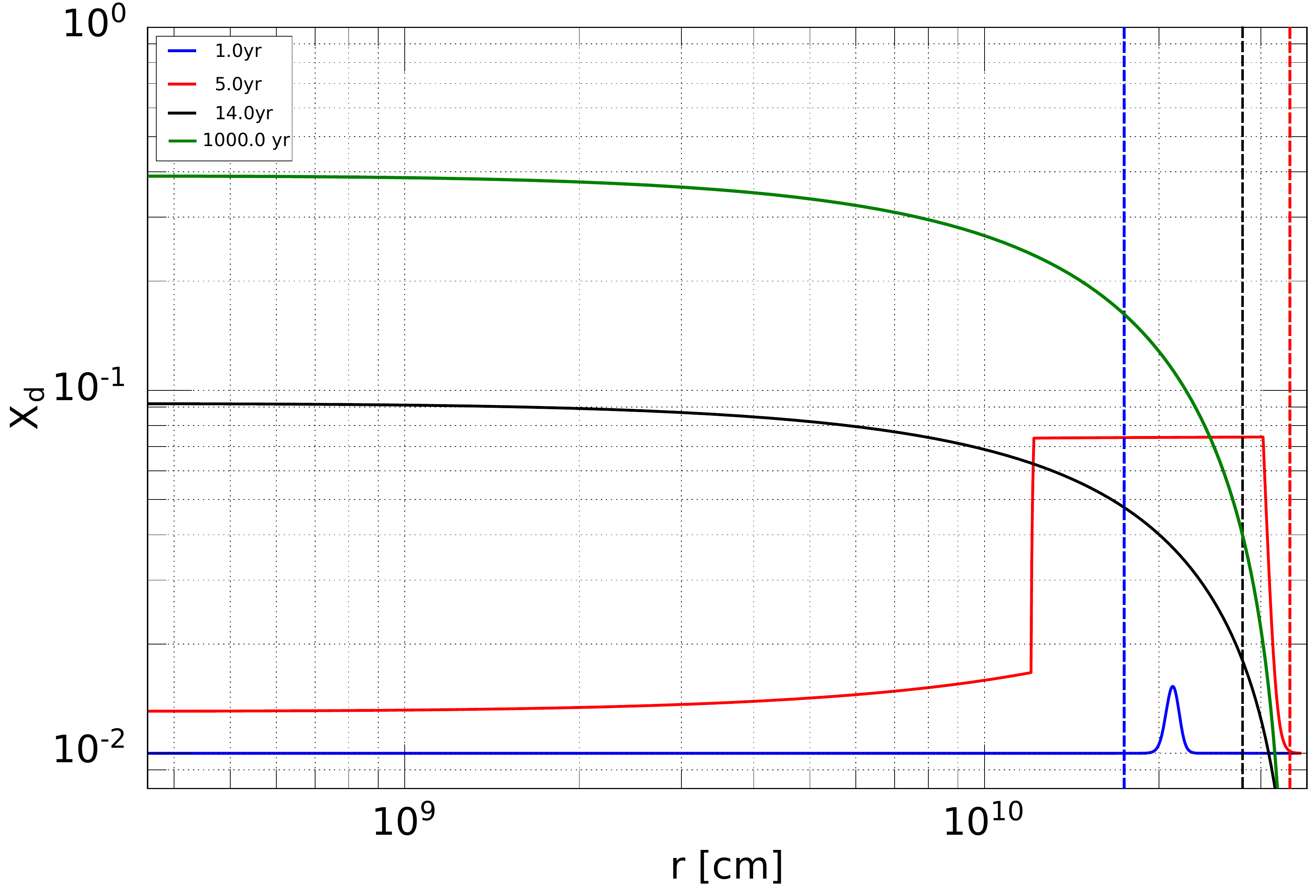}
   \caption{Temporal and radial evolution of the dust mass fraction $X_{\rm d}$ in the high-luminosity model
   ($\dot M_p = 10^{-6}\,M_\oplus$ yr$^{-1}$, $\dot M_{\rm plan} = 0$)
   at 1, 5, 14, and $10^3$ yr from the start of the simulation.  Vertical lines of matched color
   mark the position of the RCB.   Simulation is initiated with $X_d = 0.01$.  Pebble accretion injects
   dust in the outer radiative layer, producing an abundance bump that grows, shifts inward due to dust
   settling, and is flattened by a local convective instability.  As settling dust reaches the inner 
   convection zone, the RCB shifts outward to the Bondi radius.  A cycle is initiated involving 
   dust ejection by convection that then forms a temporary, thin radiative layer just inside the outer
   computational boundary (Figure \ref{fig:main}).
   The equilibrium $X_d$ in the envelope continues to rise until the model
   reaches an equilibrium configuration after $\sim 10^3$ yr.}
    \label{fig:xdust}
\end{figure}

\subsection{Evolution of the Envelope Mass and Metallicity in the Baseline Model}\label{s:evol}

The growth of the dust mass embedded in the envelope is shown in Figures \ref{fig:massratio} and
\ref{fig:dustnew} for the default case of icy pebbles.   We present a sample of
pebble accretion rates, extending between $10^{-7}$ and $10^{-6}\,M_\oplus$ yr$^{-1}$.
Particles moving near the fragmentation speed settle rapidly through the outer radiative layer.  
At the two highest accretion rates, $\dot M_p = (0.3,1)\times 10^{-6} M_\oplus$ yr$^{-1}$, the envelope 
is saturated with dust after a few thousand years, and the H$_2$ mass also saturates
(red curves in Figure \ref{fig:dustnew}).  Beyond this point, convection continuously ejects most of the 
accreted solids back into the disk in the form of small grains. 
That is,
\begin{equation}\label{eq:xdust1}
    \left(X_d \cdot 4\pi r^2 \rho V_{\rm con}\right)_{r = R_{\rm B}} \sim \dot M_p.
\end{equation}

On the other hand, in the most slowly accreting model ($\dot M_p = 1\times 10^{-7} M_\oplus$ yr$^{-1}$), 
the core mass continues to
grow at the imposed pebble accretion rate.  The masses of dust and gas in the envelope still reach a steady state,
showing an almost identical gas/dust ratio near unity (blue curves in Figures \ref{fig:massratio} and \ref{fig:dustnew}).

Here we have made the default assumption that the core-envelope boundary is transmitting to silicate dust.
Alternatively, when sublimation of silicate dust does impose a strong barrier to core growth, a very different
mass partitioning between core and envelope is obtained.
It is the dust mass stored in the envelope which grows as pebbles continue to accumulate; the core
mass remains frozen (cyan curves in Figure \ref{fig:dustnew}).  

The preceding results, which are summarized in Table \ref{tbind}, can be readily understood in terms of the evolving position $R_{\rm rad-con}$ of the outer radiative-convective boundary (RCB), as shown in Figure \ref{fig:main}. 
At the higher pebble accretion rates, the bound envelope begins with a significant radiative
layer but quickly captures pebble material.  The dust created in this outer layer rapidly settles toward the
interior, and the RCB pushes rapidly outward in reseponse to the growing opacity, until convection extends past the Bondi radius. 
By contrast, in the lowest-luminosity ($\dot M_p = 10^{-7}\,M_\oplus$ yr$^{-1}$) case the RCB never approaches 
the Bondi radius.

Recycling of pebble debris sets in at lower $\dot M_p$ if a modest fraction of the solid accretion flux is carried by large, penetrating planetesimals.
  For example, the $\dot M_p = 10^{-7}\,M_\oplus$ yr$^{-1}$ model becomes fully convective after adding $\dot M_{\rm plan} = 10^{-7}\,M_\oplus$ yr$^{-1}$
  in planetesimal accretion heating (Table \ref{tbind}).

\begin{table}
\begin{center}
\caption{Derived Parameters of the Model}
{\begin{tabular}{lcccccr}

\hline
\noalign{\smallskip}
 $\dot{M}_{\rm p}$ [$M_\oplus$ yr$^{-1}$] & M$_{\rm H_2}^{\rm sat}$ [$M_\oplus$] & M$_{\rm Z}^{\rm sat}$ [$M_\oplus$] & $E_{\rm bind}^{\rm sat}$ [erg] &$\dot{M}_{\rm sett,core}/\dot{M}_{\rm p}$& R$_{\rm RCB}$/R$_{\rm B}$ \\
\hline

$10^{-7}$ &3.28$\times 10^{-6}$&2.05$\times 10^{-6}$ & 6.65$\times 10^{33}$ &1& 0.45 \\

$1.5\times10^{-7}$ &3.58$\times 10^{-6}$&2.54$\times 10^{-6}$ & 7.63$\times 10^{33}$ &0.93 &$>1$  \\

$2\times10^{-7}$ &3.72$\times 10^{-6}$&2.75$\times 10^{-6}$ & 8.07$\times 10^{33}$ & 0.79&  $>1$  \\

$3\times10^{-7}$ &3.67$\times 10^{-6}$&2.68$\times 10^{-6}$ &7.92$\times 10^{33}$& 0.48 &$>1$\\

$10^{-6}$ &3.48$\times 10^{-6}$& 2.42$\times 10^{-6}$ &7.36$\times 10^{33}$& 0.11 &$>1$  \\

$10^{-5}$ &3.41$\times 10^{-6}$& 2.32$\times 10^{-6}$ &7.14$\times 10^{33}$& 0.013 &$>1$  \\

$10^{-7}$ ($\dot{M}_{\rm plan}=10^{-7}$) &2.72$\times 10^{-6}$& 1.44$\times 10^{-6}$ &5.19$\times 10^{33}$& 0.61 &$>1$  \\

$10^{-6}$ ($\dot{M}_{\rm plan}=10^{-7}$) &2.42$\times 10^{-6}$& 1.09$\times 10^{-6}$ &4.37$\times 10^{33}$& 0.072 &$>1$  \\

\hline	
\end{tabular}}
\label{tbind}
\end{center}
\tablecomments{Quantities are evaluated after the accretion of gas and solids into the envelope has saturated due to convective reflux, and the envelope profile has stabilized. 
Column 3:  total mass of H$_2$/He.  Column 4:  total mass of ice, water vapor, and silicates.  Column 5:  binding
energy of the envelope (volume integral of $\rho(e-GM(r)/r)$), where $e$ is internal energy.
}
\end{table}

One observes in the high-luminosity models with refluxing a rapid limit cycle involving
the transient appearance of a narrow radiative layer just interior to the computational boundary (Figure \ref{fig:main}).
This effect is driven
by the finite lifetime of the convective eddies:  the accretion of low-metallicity gas pushes the RCB inward, but
this material then mixes with the dusty envelope, forcing the convection back outward. This cycle is shown in more
detailed in Figure \ref{fig:xdust}, which plots the spatial and temporal evolution of the dust mass fraction. 
Dust deposited in the outer radiative envelope due to sandblasting settles inward quickly, pushing the RCB outwards. 
When the envelope becomes fully convective, it loses dust in the outer scale height or so, which creates
a transient radiative zone.  As this cycle continues, the dust mass builds up in the envelope;
the radial width of the transient radiative zone decreases with time, until the envelope reaches a steady state 
around $t \sim 10^3$ yr.

This effect is, to some extent, an artifact of a one-dimensional treatment of the convection, combined with
the presence of a sharp outer computational boundary.   More realistically, the dust abundance is significantly
larger in upflowing convective plumes than in gas accreted from beyond the Bondi radius, with the latter 
providing opacity holes through which much of the radiation escapes.  One can expect a more gradual 
transition in $X_d$ when the surrounding disk is included in the computational model.

\subsection{Dust Size Distribution}

Details of the dust size distribution have already been presented in Figures \ref{fig:psizer} and
\ref{fig:dustsize}.  The dust size is limited by fragmentation, either resulting from turbulent
stirring (Equation (\ref{eq:frag1})) or from secular drift (Equation (\ref{eq:frag2})), with the 
drag law (Epstein or Stokes) being determined self-consistently.  Figure \ref{fig:psizer} shows that
turbulent fragmentation dominates near the Bondi radius, but the secular drift speed is higher
inside $\sim 10^{10}$ cm from the core.   A substantially larger dust size
also results near the Bondi radius in cases where the outer envelope is radiative.

\vskip .3in
\section{Summary and Discussion}
\label{conc}

We have identified a significant constraint on the growth of planetary cores by the
accretion of small (mm-cm sized) pebbles.   When ambient conditions are optimal for
pebble accretion, pebbles are susceptible to fragmentation (Section \ref{s:estimates}).
Pebble accretion is most efficient when the drift speed of a pebble through the embedding
gas toward the local center of gravity is comparable to flow speed of the gas near the
the core.  In such a circumstance, the pebble drift speed far exceeds
the fragmentation speed of an amorphous
conglomerate of (sub)-micron sized grains.  The water ice component of pebbles
  is rapidly lost by sublimation, releasing embedded silicate grains and preventing
  the water ice from reaching the core.

A high pebble accretion rate, 
as needed to fuel the rapid growth of a small core into a Neptune-mass planet 
($t_{{\rm acc},p} = M_c/\dot M_p \lesssim 3\times 10^5$ yr), implies (i) a significant dust mass
fraction in the medium surrounding the core, and (ii) efficient binary collisions 
between pebbles and, even more so, between pebbles and dust grains suspended in the gas.
The build-up of small particles in the envelope, which do not accrete efficiently onto
  the core, triggers rapid sandblasting of high-speed pebbles around the Bondi radius.
    Indeed, convection sets in within the outer envelope only when the mass fraction of
  particles smaller than $\lambda_{\rm max}(T_{\rm disk}) \sim 30\, \mu$m is larger than $\sim 0.3$.
  
Pebble fragmentation does not, by itself, imply inefficient core growth.
Even as their density exponentiates in the outer radiative layer,
dust particles moving near the fragmentation speed will settle fairly rapidly
reaching the inner convective layer in a year or less.
When the imposed pebble accretion rate exceeds $\sim 10^{-7}\,M_\oplus$ yr$^{-1}$,
we find that convective motions grow and extend
beyond the planet's gravitational sphere of influence as defined by
ambient disk conditions (Section \ref{s:results}).  Then some of the debris of the accreted pebbles is
recycled back into the disk, and the accretion rate of the remaining debris onto 
a $0.3\,M_\oplus$ core stabilizes at $\sim 1\times 10^{-7}\,M_\oplus$ yr$^{-1}$, corresponding
to an e-folding time of $\sim 3$ Myr. The loss of dust and gas eventually counterbalances accretion, 
causing the solid mass in the envelope to plateau. 
The envelope density profile remains shallow and its entropy equilibrates
near the ambient disk entropy.  

When the pebbles are icy, the water component sublimates in the outer envelope and does
not accrete onto the core.  Ice sublimation extracts heat from the gas in the outer 
envelope, which raises the inner density of the envelope.  We find, however, that continuing
accretion of hydrogen from the disk limits the water vapor mass fraction and maintains
a signficant H$_2$/He component in the envelope.  This in turn limits the density growth
in the inner envelope that is caused by heat extraction from the outer envelope.

Embedded grains do settle toward the
core through the inner convective envelope.  The rate at which the core gains silicate
mass is estimated from the radial drift time (Equation (\ref{eq:mdotsett})).  This accretion
rate $\dot M_{\rm sett}$ is proportional to the dust concentration in the inner envelope.  Allowing for a relative
concentration $f_{\rm turb}$ of dust at the base of the envelope as compared to its interior,
$\dot M_{\rm sett}$ is less than $10\,f_{\rm turb}\%$ of the imposed pebble accretion rate,
for $\dot M_p = 10^{-6}\,M_\oplus$ yr$^{-1}$.  By comparision, a short accretion time $M_c/\dot M_p \sim
3\times 10^5$ yr needed to build a Neptune-mass planet in the lifetime of the protoplanetary disk.

The aforementioned effects were demonstrated using a self-consistent, spherical model of the bound envelope
(Section \ref{s:model})
which includes the thermodynamics of water phase exchange, pebble destruction, dust collision and size evolution, 
convection in the mixing length approximation, multiple luminosity sources, dust transport handled using
the advection-diffusion equation, analytical prescriptions for dust and gas radiative opacity, and 
tabulated non-ideal EOSs for water and hydrogen.   The steady steady behavior produced by this model
is in agreement with the convective/radiative structure seen in a steady-state spherical model (Section
\ref{s:expand}). 


The specific example studied numerically is a core of mass $0.3\,M_\oplus$.  A simple
analytic model of a hydrostatic envelope loaded with dust gives a nearly identical
result (Equation (\ref{eq:xdmin})) for the critical deep heating rate (due to planetesimal
accretion) above which refluxing balances pebble accretion.
The growth time of the core due to the settling of pebble debris in the inner
  envelope scales as $t_{{\rm acc},c} = M_c/\dot M_{\rm sett}(R_c) \propto
  (X_{d,\rm env}\rho_{g,\rm disk})^{-1} M_c^{-4/3}$;  here $X_{d,\rm env}$ is the envelope
  metallicity and in Equation (\ref{eq:mdotsett}) we assume that the inward drift speed of the pebble
  debris is the fragmentation speed $V_f$.  Given a fixed ratio of
pebble drift speed to ambient gas sound speed ($\rho_g \propto M_c^{-1}$ at the Bondi
radius, from Equation (\ref{eq:rhodisk})), refluxing becomes slightly more difficult as
$M_c$ increases, $t_{{\rm acc},c}  \propto X_{d,\rm env}^{-1}M_c^{-1/3}$.
Setting aside this constraint of efficient pebble accretion, the minimum $X_{d,\rm env}$
  that forces convection at the Bondi radius scales  as 
  $X_{d,\rm env} \propto \rho_{g,\rm disk}^{-1}M_c^{-2/3}\,t_{{\rm acc},c}$ (Equation (\ref{eq:xdmin})).
  Hence, the core grows moderately faster at the onset of refluxing as its mass
  increases, $t_{{\rm acc},c} \propto M_c^{-1/3}$ (the dependence on $\rho_{g,\rm disk}$ cancels).
  Finally, we
  may consider the minimum $X_{d,\rm env}$ needed for dust refluxing across the Bondi radius to balance pebble
accretion, $X_{d,\rm env} \rho_g V_{\rm con} 4\pi R_{\rm B}^2 \sim M_c/t_{{\rm acc},p}$,
where $t_{{\rm acc},p}$ is the accretion time associated with the imposed pebble trapping rate.
Given the preceding scaling for $t_{{\rm acc},c}$, one finds from Equation (\ref{eq:mach})
that $V_{\rm con} \propto M_c^0 \rho_{g,\rm disk}^{-1/3}$.  Hence pebble accretion is
cancelled for $X_{d,\rm env} \propto \rho_{g,\rm disk}^{-2/3} M_c^{-1}t_{{\rm acc},p}^{-1}$,
which once again decreases with increasing $M_c$.

\subsection{Caveats}

1.  An important, unresolved issue about pebble accretion is whether it
proceeds mainly from a thin, pebble-dominated disk \citep{goldreich2004b}, 
or by trapping from a gas-dominated medium \citep{ormelklahr}.
Our calculations address the latter possibility,
but may still be applicable to the first situation when the bound envelope is
convective and mixes embedded solids thoroughly in all three dimensions.

2. The 3D numerical simulations of \cite{popovas} follow the trajectories of solid
particles through the gas around a planetary core embedded in a shearing gas disk.
Allowing for heating of the gas by transfer of accretion energy from the pebbles, they
find evidence for extended cold downflows, which help to advect pebbles to the core.
However, they also find that particles drifting at the silicate fragmentation 
speed of $\sim 1$~m~s$^{-1}$ can be cycled repeatedly within convective flows;
more generally, the dynamics of small particles with short stopping times is not
resolved near the core.  Our approach to dust accretion suggests that such particles
experience slow secular drift toward the core, otherwise being entrained in the
rapid convective flows and surviving for a long time in the envelope.  

3. The main limitation of our model is the use of a spherical approximation
and the neglect of rotational effects near the Bondi radius.  The physical
processes we investigate need to be incorporated into a global, 3D simulation
of a core embedded in a gas disk.  Efficient pebble accretion without strong fragmentation might  
be easier if the core were surrounded by 
an orbiting gas disk that trapped pebbles for many orbits.  Then tightly-coupled pebbles might spiral
into the core at reduced speeds relative to the gas.  The extent to which such centrifugally-supported
structures form around low-mass cores (Bondi radius $\ll$ Hill radius) remains an open question.

4. The recycling of ejected debris back to the disk may enhance the 
metallicity of the accreted material, but we have argued that this does
not significantly perturb the equilibrium envelope state as long as
the ejected debris is diluted by mixing with disk gas.

5. The effect of silicate dust sublimation in the inner envelope was not included
in our default envelope model, and is not needed to suppress core growth.
Nonetheless, we argue that the re-evaporation of silicate ``rain''  can 
have a significant additional suppressing effect on solid accretion by the core
when its mass exceeds $0.2-0.3\,M_\oplus$.

6. The size of a grain near the threshold for fragmentation is typically much
smaller than the assumed pebble size of 1 mm, and is small enough
to produce significant opacity ($a_d < \lambda_{\rm max}(T)$).  Furthermore, we have 
argued that reassembly of spallated monomers into smaller amorphous conglomerates
cannot happen fast enough to limit the exponential growth of opacity
by the sandblasting process.  Nonetheless, the effect of collisions of pebbles with
small, marginally-fragmenting conglomerates needs further attention.

\subsection{Implications and Future Directions}

1. Pebble destruction by sandblasting \citep{schrap1, jacquet} may be important
well outside the quasi-spherical, bound envelope treated here.  Hydrodynamic models of pebbles and gas interacting
with a core should take into account the feedback of dust-enriched gas back into the disk, and its effect on radiation
transfer and gas entropy. 

2. When the ambient disk temperature lies not too far below the sublimation temperature of water ice, we find
a strong gradient in gas density near the Bondi radius, associated with the absorption of heat by sublimating ice.
This means that the gas envelope accreted around a core can maintain a well-defined boundary
even while the envelope remains thermodynamically coupled to the disk, that is, while it
maintains a similar specific entropy.

3. Our models with a reduced luminosity  -- which fail to develop convection near the
Bondi radius -- maintain only a modest metallicity in the envelope in spite of rapid pebble accretion, 
typically $\sim 20$ times Solar. That is
because the accretion of pebbles is accompanied by the continued accretion of hydrogen and helium, as sublimating ice
absorbs heat from the envelope and the gas component contracts and cools.  
This result, if verified at large core masses, would argue against the formation of icy Neptune-mass
planets by pebble accretion from a gas-rich disk.  There is also a synergy between this result and recent accretion
models of Jovian planets, which never develop a substantial core
(see, e.g., \citealt{helled}); and recent
measurements of the gravitational moments of Jupiter,
which allow for a dilute core in that planet \citep{wahl}.

4. We have not considered processes that might mix core material back into the envelope.
In the generic situation considered here, the envelope has a higher entropy and lower metallicity than the
core, meaning that the core-envelope boundary is not susceptible to either salt-fingering (higher metallicity
and higher entropy on the outside) or semi-convection (lower metallicity and lower entropy on the outside).
However, convective dredge-up could supplement the process of dust loading of a gas envelope by pebble accretion.

5. Ongoing pebble accretion has important implications for the retention of planetary atmospheres during
a giant impact phase.  The saturation in the growth of the envelope
implies a strong limit to the gravitational binding energy of the envelope
to the core (Table \ref{t4}).  Such a low-mass envelope is more susceptible to expulsion by giant impacts.

6. This work suggests an interesting scenario for the origin of high temperature minerals (crystalline silicates and 
CAIs) observed in comets (for example \cite{hanner1,hanner2,brownlee}).  In our model, pebbles are processed 
thermally and mechanically at very high temperatures in the inner envelope before being convectively transported 
and ejected back into the disk. The combination of high temperature and high envelope metal enrichment satisfies 
some of the basic constraints on the birth environment of  these minerals;  we leave detailed modeling to future work.

The broad lesson here is that the structure of planetary atmospheres during the planet assembly phase can be dramatically
altered by pebble accretion, and more generally is very sensitive to the mode of solid delivery.   
In the context of future 3D accretion simulations and time-dependent envelope models, our work implies
that it is essential to incorporate a local fragmentation model for embedded grains and to treat their
evolution separately from infalling pebbles.  The combination of accretion heating by settling solids
with the build-up of grain opacity is found to have a powerful feedback effect on gas flows in the outer envelope.
By contrast, many current planet formation models allow for the growth of grains that are sourced
by infalling planetesimals, while ignoring fragmentation
feedback on the grain size and therefore on the mass fraction and opacity of settling grains (e.g. \citealt{mordaop, ormel}).
It is also essential to simulate the interaction of lower core masses
with pebble-loaded gas disks, as our model shows strong feedback effects
even for $M_c = 0.3\,M_\oplus$, and implies a significant modification of the 
solid particle population in the ambient disk through refluxing.

\acknowledgements
M.A.-D. thanks the Department of Physics at the American University of Beirut where he spent one semester as a visiting faculty. 
The authors thank T. Birnstiel for discussions on dust size evolution, D. Valencia for discussions on opacities and 
equations of state, and K. Menou for discussions that helped to motivate this project.  Finally, we thank the referee
Anders Johansen for comments on our treatment of collisional fragmentation.

\appendix
\section{Small-grain Dust Opacity}\label{s:kapgr}

Here we motivate our choice of Rosseland mean opacity $\kappa_{d,\rm R}(T)$ for small grains (of radius $a_d \ll 
\lambda_{\rm max}(T) = hc/4.95\,kT$.  The grain opacity used in our atmosphere models is the minimum of this value and the
geometric opacity $\kappa_{\rm geom}$ (see Equation (\ref{eq:kapgr})).  Given a wavelength dependence $\kappa(a_d,\lambda) = 
Q' \cdot (2\pi a_d/\lambda) \cdot \kappa_{\rm geom}(a_d) = Q'\cdot 3\pi X_d / 2\rho_s\lambda$ for grains of density $\rho_s$, 
the Rosseland mean works out to 
\begin{equation}
\kappa_{d,\rm R} = {\int d\nu\, dB_\nu/dT \over \int d\nu \kappa_d^{-1} dB_\nu/dT}
= 3.6 \,\kappa_d\left(a_d,\lambda = {hc\over kT}\right) = {3.6\over 4.95}\kappa_d(a_d,\lambda_{\rm max}).
\end{equation}
Here $\nu = c/\lambda$ and $B_\nu(\nu,T)$ is the Planck function.  The opacity of small 
silicate-carbon grains, of density $\simeq 2.5$ g cm$^{-3}$ is derived by
\cite{ossen} to be $\kappa_d = 7.0$ cm$^2$ g$^{-1}$ at $\lambda = 1.3$ mm (see their Figure 9).
This corresponds to $Q' = 0.50$ and a Rosseland mean 
\begin{equation}\label{eq:kapR2}
\kappa_{d,\rm R}(T) = 230\,X_d\left({T\over 100~{\rm K}}\right)\quad {\rm cm^2~g^{-1}},
\end{equation}
as weighted by gas mass.  Equivalently, $Q_R(a_d,T) = 0.73\, Q' \cdot 2\pi a_d/\lambda_{\rm max}(T) = 
0.36\,\cdot 2\pi a_d/\lambda_{\rm max}(T)$.  

Ice coatings deposited by condensation onto silicate grain surfaces could reduce the opacity
by a factor $1/(6-7)$ (Figure 9 of \cite{ossen}).

\begin{figure}
\epsscale{0.75}
\plotone{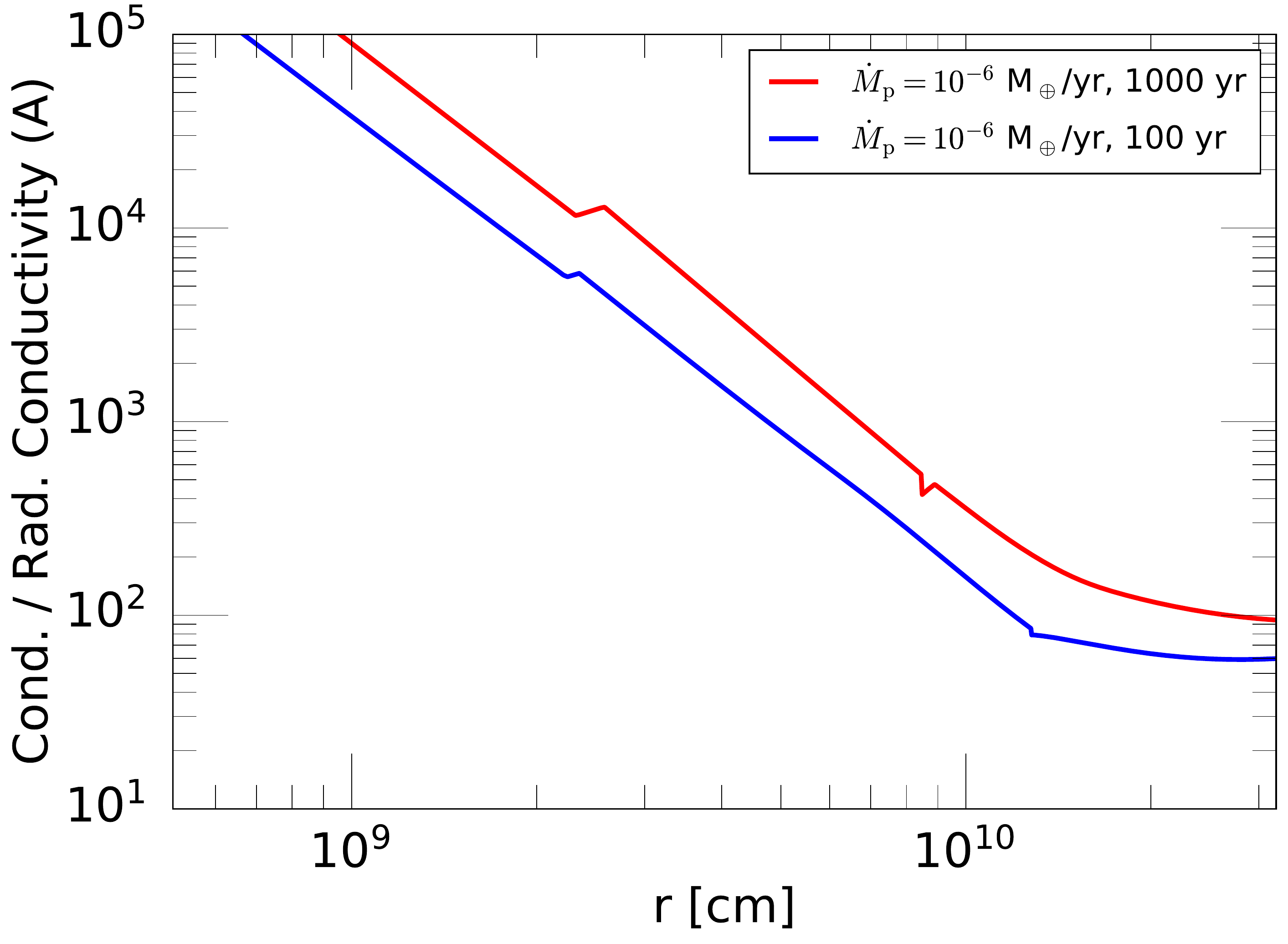}
   \caption{Ratio $A$ of convection and radiative conductivities, as defined in Equation (\ref{cond}),
   in the high-luminosity model at two different times.}
    \label{fig:conveff}
\end{figure}

\begin{figure}
\epsscale{0.75}
\plotone{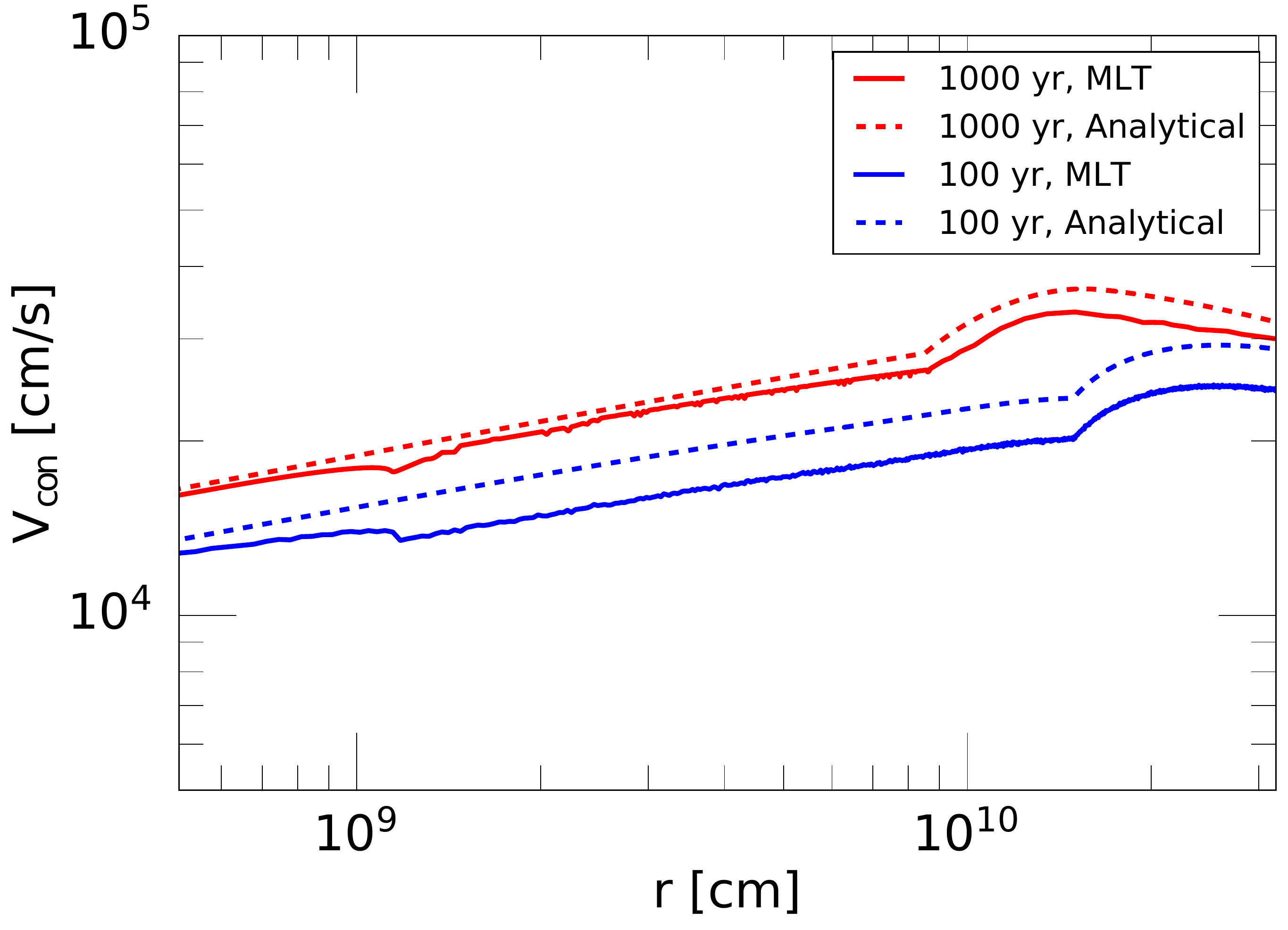}
\caption{The convective speed at $t = 10^2$ and $10^3$ yr in the high-luminosity model, as derived from the mixing-length approximation
  (Equation (\ref{eq:dtot}); solid lines), and from the analytic estimate (Equation (\ref{eq:lcore}); dashed lines).  A larger choice of mixing length (e.g. $\alpha = 2-3$) would bring these
   curves even closer together.}
    \label{fig:vcon}
\end{figure}

\section{Convective Efficiency}\label{s:coneff}

 The relative importance of convection and radiation in transporting energy through the envelope
is encapsulated in the parameter $A$ given by Equation (\ref{cond}).  This is plotted as a function of
radius in Figure \ref{fig:conveff} for our highest $\dot M_p$ model, in which convection
  reaches the outer computational boundary and envelope growth is quenched.  In both cases, 
$A$ attains a minimum value $O(10^2)$ in the outer envelope, corresponding to energy transport dominated
by convection.  

Figure \ref{fig:vcon} shows that the simple approximation $F_{\rm con} \simeq \rho V_{\rm con}^3$ is adequate
in the convective regions of the envelope after it has reached a steady state.
In the same formulation of mixing length theory used in this paper, one finds $F_{\rm con} \simeq 
0.36 \alpha \rho V_{\rm con}^3$, where $\alpha$ is the ratio of mixing length to pressure scale height.
We take $\alpha = 1$ in the numerical model.  

In our high-luminosity models, the superadiabaticity $\nabla - \nabla_{\rm ad} \sim 8\times 10^{-6}$
near the core, comparable to the value in the interiors of giant planets, and implying almost perfectly
adiabatic convection.  This quantity rises to $\sim 10^{-3}$ near the Bondi radius, where the assumption 
of full adiabaticity approaches the limit of applicability.


\begin{thebibliography}{00}

\bibitem[\protect\citeauthoryear{Alibert}{2017}]{alibert} Alibert Y., 2017, A\&A, 606, A69

\bibitem[Beitz et al.(2011)]{dustnew1} Beitz, E., G{\"u}ttler, C., Blum, J., et al.\ 2011, \apj, 736, 34

\bibitem[\protect\citeauthoryear{Birnstiel, Klahr, \& Ercolano}{2012}]{til} Birnstiel T., Klahr H., Ercolano B., 2012, A\&A, 539, A148 

\bibitem[Blum \& Wurm(2008)]{blum} Blum, J., \& Wurm, G.\ 2008, \araa, 46, 21

\bibitem[Brouwers et al.(2018)]{Brouwers} Brouwers, M.~G., Vazan, A., \& Ormel, C.~W.\ 2018, \aap, 611, A65 

\bibitem[\protect\citeauthoryear{Brownlee et al.}{2006}]{brownlee} Brownlee D., et al., 2006, Sci, 314, 1711 


\bibitem[Chambers(2017)]{chambers2017} Chambers, J.\ 2017, \apj, 849, 30 

\bibitem[Chiang \& Goldreich(1997)]{cg1997} Chiang, E.~I., \& Goldreich, P.\ 1997, \apj, 490, 368

\bibitem[\protect\citeauthoryear{Cox \& Giuli}{1968}]{cox} Cox J.~P., Giuli R.~T., 1968, Principles of stellar structure  





\bibitem[Fontaine et al.(1977)]{fontaine} Fontaine, G., Graboske, H.~C., Jr., \& van Horn, H.~M.\ 1977, \apjs, 35, 293 

\bibitem[\protect\citeauthoryear{Freedman et al.}{2014}]{freedman} Freedman R.~S., Lustig-Yaeger J., Fortney J.~J., Lupu R.~E., Marley M.~S., Lodders K., 2014, ApJS, 214, 25 
















\bibitem[Goldreich et al.(2004a)]{goldreich2004} Goldreich, P., Lithwick, Y., \& Sari, R.\ 2004, \araa, 42, 549 

\bibitem[Goldreich et al.(2004b)]{goldreich2004b} Goldreich, P., Lithwick, Y., \& Sari, R.\ 2004, \apj, 614, 497 

\bibitem[G{\"u}ttler et al.(2010)]{guttler10} G{\"u}ttler, C., Blum, J., Zsom, A., et al.\ 2010, \aap, 513, A56

\bibitem[Guyer et al.(2009)]{guyer} Guyer, J.~E., Wheeler, D., \& Warren, J.~A.\ 2009, Computing in Science and Engineering, 11, 6 

\bibitem[\protect\citeauthoryear{Hanner et al.}{1994}]{hanner1} Hanner M.~S., Hackwell J.~A., Russell R.~W., Lynch D.~K., 1994, Icar, 112, 490 

\bibitem[\protect\citeauthoryear{Hanner, Lynch, \& Russell}{1994}]{hanner2} Hanner M.~S., Lynch D.~K., Russell R.~W., 1994, ApJ, 425, 274 

\bibitem[Hayashi(1981)]{hayashi} Hayashi, C.\ 1981, Progress of Theoretical Physics Supplement, 70, 35 

\bibitem[Haynes et al.(1992)]{haynes} Haynes, D.~R., Tro, N.~J., \& George, S.~M.\ 1992, Journal of Physical Chemistry, 96, 8502 

\bibitem[Harvey \& Lemmon(2013)]{nist} Harvey, Allan H., \& Lemmon, Eric W. 2013, NIST Standard Reference Database 10, June 03

\bibitem[Hill et al.(2015)]{dustnew3} Hill, C.~R., Hei{\ss}elmann, D., Blum, J., et al.\ 2015, \aap, 573, A49


\bibitem[Hori \& Ikoma(2011)]{hori2011} Hori, Y., \& Ikoma, M.\ 2011, \mnras, 416, 1419 

\bibitem[Hueso \& Guillot(2005)]{hueso} Hueso, R., \& Guillot, T.\ 2005, \aap, 442, 703 

\bibitem[Ingersoll(1969)]{ingersoll69} Ingersoll, A.~P.\ 1969, Journal of Atmospheric Sciences, 26, 1191


\bibitem[\protect\citeauthoryear{Jacquet \& Thompson}{2014}]{jacquet} Jacquet E., Thompson C., 2014, ApJ, 797, 30 

\bibitem[Jessberger \& Kissel(1991)]{halley} Jessberger, E.~K., \& Kissel, J.\ 1991, IAU Colloq.~116: Comets in the post-Halley era, 167, 1075 

\bibitem[Johansen \& Lambrechts(2017)]{jonlam} Johansen, A., \& Lambrechts, M.\ 2017, Annual Review of Earth and Planetary Sciences, 45, 359 



\bibitem[Kippenhahn et al.(2012)]{kippen} Kippenhahn, R., Weigert, A., \& Weiss, A.\ 2012, Stellar Structure and Evolution: , Astronomy and Astrophysics Library.~ISBN 978-3-642-30255-8.~Springer-Verlag Berlin Heidelberg, 2012,  

\bibitem[Krieger (1967)]{Krieger} Krieger, F. J. 1967, Rand Corp. Memo. RM-5337-PR.

\bibitem[Krijt et al.(2015)]{krijt} Krijt, S., Ormel, C.~W., Dominik, C., \& Tielens, A.~G.~G.~M.\ 2015, \aap, 574, A83 

\bibitem[Kurokawa \& Tanigawa(2018)]{kurokawa} Kurokawa, H., \& Tanigawa, T.\ 2018, \mnras, 479, 635 

\bibitem[Kuwahara et al.(2019)]{kuwahara} Kuwahara, A., Kurokawa, H., \& Ida, S.\ 2019, arXiv:1901.08253 


\bibitem[\protect\citeauthoryear{Lambrechts \& Johansen}{2012}]{lamborg} Lambrechts M., Johansen A., 2012, A\&A, 544, A32 

\bibitem[\protect\citeauthoryear{Lambrechts \& Johansen}{2014}]{lamb1} Lambrechts M., Johansen A., 2014, A\&A, 572, A107

\bibitem[Lambrechts \& Lega(2017)]{lega} Lambrechts, M., \& Lega, E.\ 2017, \aap, 606, A146 


\bibitem[Leconte \& Chabrier(2012)]{leconte} Leconte, J., \& Chabrier, G.\ 2012, \aap, 540, A20 

\bibitem[Lee \& Chiang(2015)]{lee} Lee, E.~J., \& Chiang, E.\ 2015, \apj, 811, 41 


\bibitem[Lifshitz \& Pitaevskii(1981)]{lp1981} Lifshitz, E.~M., \& Pitaevskii, L.~P.\ 1981, Course of theoretical physics

\bibitem[Lozovsky et al.(2017)]{helled} Lozovsky, M., Helled, R., Rosenberg, E.~D., \& Bodenheimer, P.\ 2017, \apj, 836, 227 

\bibitem[Mordasini(2014)]{mordaop} Mordasini, C.\ 2014, \aap, 572, A118

\bibitem[Militzer \& Hubbard(2013)]{milit1} Militzer, B., \& Hubbard, W.~B.\ 2013, \apj, 774, 148

\bibitem[Militzer(2013)]{milit2} Militzer, B.\ 2013, \prb, 87, 014202

\bibitem[Okuzumi et al.(2012)]{porosity} Okuzumi, S., Tanaka, H., Kobayashi, H., \& Wada, K.\ 2012, \apj, 752, 106 

\bibitem[\protect\citeauthoryear{Ormel \& Klahr}{2010}]{ormelklahr} Ormel C.~W., Klahr H.~H., 2010, A\&A, 520, A43 

\bibitem[\protect\citeauthoryear{Ormel}{2014}]{ormel} Ormel C.~W., 2014, ApJ, 789, L18

\bibitem[\protect\citeauthoryear{Ormel, Shi, \& Kuiper}{2015}]{ormelcy} Ormel C.~W., Shi J.-M., Kuiper R., 2015, MNRAS, 447, 3512 

\bibitem[\protect\citeauthoryear{Ossenkopf \& Henning}{1994}]{ossen} Ossenkopf V., Henning T., 1994, A\&A, 291, 943 


\bibitem[\protect\citeauthoryear{Podolak, Pollack, \& Reynolds}{1988}]{podolak} Podolak M., Pollack J.~B., Reynolds R.~T., 1988, Icar, 73, 163 


\bibitem[\protect\citeauthoryear{Pollack et al.}{1986}]{pollack1986} Pollack J.~B., Podolak M., Bodenheimer P., Christofferson B., 1986, Icar, 67, 409 


\bibitem[Popovas et al.(2018)]{popovas} Popovas, A., Nordlund, A., Ramsey, Jon P., Ormel, Chris W. 2018, \mnras, 479, 5136

\bibitem[Popovas et al.(2019)]{popovas19} Popovas, A., Nordlund, {\r{A}}., \& Ramsey, J.~P.\ 2019, \mnras, 482, L107
  

\bibitem[Rafikov(2006)]{rafikov2006} Rafikov, R.~R.\ 2006, \apj, 648, 666

\bibitem[Pruppacher \& Rasmussen(1979)]{pruppacher79} Pruppacher, H.~R., \& Rasmussen, R.\ 1979, Journal of Atmospheric Sciences, 36, 1255




\bibitem[Schr{\"a}pler \& Blum(2011)]{schrap1} Schr{\"a}pler, R., \& Blum, J.\ 2011, \apj, 734, 108

\bibitem[Schr{\"a}pler et al.(2012)]{dustnew2} Schr{\"a}pler, R., Blum, J., Seizinger, A., et al.\ 2012, \apj, 758, 35

\bibitem[Schr{\"a}pler et al.(2018)]{schrap2} Schr{\"a}pler, R., Blum, J., Krijt, S., \& Raabe, J.-H.\ 2018, \apj, 853, 74 

\bibitem[Stevenson(1984)]{stevenson84} Stevenson, D.~J.\ 1984, Lunar and Planetary Science Conference, 15, 822 

\bibitem[Th{\'e}ado \& Vauclair(2012)]{theado} Th{\'e}ado, S., \& Vauclair, S.\ 2012, \apj, 744, 123 

\bibitem[Thompson \& Stevenson(1988)]{ts1988} Thompson, C., \& Stevenson, D.~J.\ 1988, \apj, 333, 452

\bibitem[Thoul et al.(1994)]{thoul} Thoul, A.~A., Bahcall, J.~N., \& Loeb, A.\ 1994, \apj, 421, 828 




\bibitem[Vazan et al.(2016)]{vazan2016} Vazan, A., Helled, R., Podolak, M., \& Kovetz, A.\ 2016, \apj, 829, 118 

\bibitem[\protect\citeauthoryear{Venturini, Alibert, \& Benz}{2016}]{vent} Venturini J., Alibert Y., Benz W., 2016, A\&A, 596, A90 

\bibitem[Wada et al.(2008)]{wada} Wada, K., Tanaka, H., Suyama, T., Kimura, H., \& Yamamoto, T.\ 2008, \apj, 677, 1296 

\bibitem[Wada et al.(2009)]{wada2} Wada, K., Tanaka, H., Suyama, T., Kimura, H., \& Yamamoto, T.\ 2009, \apj, 702, 1490 

\bibitem[Wahl et al.(2017)]{wahl} Wahl, S.~M., Hubbard, W.~B., Militzer, B., et al.\ 2017, \grl, 44, 4649 





\end{thebibliography}
\end{document}